\documentclass[lettersize,journal]{IEEEtran}
\usepackage{amsmath,amsfonts}
\usepackage{algorithmic}
\usepackage{algorithm}
\usepackage{array}
\usepackage[caption=false,font=normalsize,labelfont=sf,textfont=sf]{subfig}
\usepackage{textcomp}
\usepackage{stfloats}
\usepackage{url}
\usepackage{verbatim}
\usepackage{graphicx}
\usepackage{cite}
\usepackage{multirow}
\usepackage{xcolor}
\usepackage{pgfplotstable}
\usepgfplotslibrary{statistics}
\pgfplotsset{grid style={dotted,gray}}
\usepackage{tikz}
\usepackage{xcolor}
\pgfplotsset{compat=1.8}
\usepackage{pgfplots}
\usetikzlibrary{spy}
\usepackage{soul}
\usepackage{lipsum}
\usepackage{fancyhdr}
\usepackage{makecell}
\usepackage{nicematrix} % <---
\begin{document}

\title{COMSPLIT: A Communication--Aware Split Learning Design for Heterogeneous IoT Platforms}

\author{Vukan Ninkovic,~\IEEEmembership{Student Member,~IEEE,} Dejan Vukobratovic, ~\IEEEmembership{Senior Member,~IEEE,} Dragisa Miskovic, ~\IEEEmembership{Member, ~IEEE,} Marco Zennaro, ~\IEEEmembership{Senior Member,~IEEE}
        % <-this % stops a space
\thanks{V. Ninkovic is with the Faculty of Technical Sciences, University of Novi Sad, Serbia, and  the Institute for Artificial Intelligence
 Research and Development of Serbia (e-mail: ninkovic@uns.ac.rs);
 D. Vukobratovic is with Faculty of Technical Sciences, University of Novi Sad,
 Serbia, (email: dejanv@uns.ac.rs);
 D. Miskovic is with the Institute for Artificial Intelligence Research and Development of Serbia (e-mail: dragisa.miskovic@ivi.ac.rs); M. Zennaro is with Science, Technology and Innovation Unit, The Abdus Salam International Centre for Theoretical Physics, Italy (email: mzennaro@ictp.it).}% <-this % stops a space
\thanks{This work has received funding from the Horizon 2020 grant agreement No 101086387; From the  Science Fund of the Republic of Serbia, grant No 6707; and from Secretariat for Higher Education and Scientific Research of the Autonomous Province of Vojvodina, grant No 142-451-3511/2023-01}}

% The paper headers
\markboth{Journal of \LaTeX\ Class Files,~Vol.~14, No.~8, August~2021}%
{Shell \MakeLowercase{\textit{et al.}}: A Sample Article Using IEEEtran.cls for IEEE Journals}

% Remember, if you use this you must call \IEEEpubidadjcol in the second
% column for its text to clear the IEEEpubid mark.

\maketitle
\setcounter{footnote}{0} 

\begin{abstract}
 %Adopting a centralized approach to neural network learning in Internet of Things (IoT) networks introduces obstacles such as increased communication overhead and concerns regarding data privacy. On the other hand, conducting learning directly at the edge device is hampered by constraints in computational resources and the necessity for adaptable approaches in fluctuating environments. 
The significance of distributed learning and inference algorithms in Internet of Things (IoT) network is growing since they flexibly distribute computation load between IoT devices and the infrastructure, enhance data privacy, and minimize latency. However, a notable challenge stems from the influence of communication channel conditions on their performance. In this work, we introduce COMSPLIT: a novel communication-aware design for split learning (SL) and inference paradigm tailored to processing time series data in IoT networks. COMSPLIT provides a versatile framework for deploying adaptable SL in IoT networks affected by diverse channel conditions. In conjunction with the integration of an early-exit strategy, and addressing IoT scenarios containing devices with heterogeneous computational capabilities, COMSPLIT represents a comprehensive design solution for communication-aware SL in IoT networks. Numerical results show superior performance of COMSPLIT compared to vanilla SL approaches (that assume ideal communication channel), demonstrating its ability to offer both design simplicity and adaptability to different channel conditions.

\end{abstract}


\section{Introduction}
\IEEEPARstart{R}{ecent} trends in the integration of Artificial Intelligence (AI) and Internet of Things (IoT) technologies represents a pivotal step in bringing inference, reasoning, and decision-making closer to the data sources. This convergence aims to optimize the utilization of resources by offloading tasks from cloud infrastructure to edge devices, thereby reducing the communication pressure on network infrastructure \cite{cao_2023}. In general, this trend leads to deployment of compute infrastructure along the entire edge-cloud continuum, facilitating faster response times for applications that cannot afford the latency inherent to computing services of cloud-based AI systems \cite{cao_2023}.

The cornerstone of the current AI revolution lies in inference algorithms based on deep neural networks, commonly known as Deep Learning (DL). Integrating DL-based inference on resource-constrained IoT edge devices holds the potential for advanced AI-enabled IoT applications. However, resource limitations of IoT devices, characterized by constrained energy, memory, and computational capabilities, pose obstacles to executing sophisticated DL models \cite{disnet_2024}. To address this problem, deployment of AI models on resource-limited IoT devices is recently addressed through distributed inference approaches \cite{khan_2021, chen_2023}. Distributed DL offers a promising avenue, enabling collaborative DL model execution across IoT devices and edge computing infrastructure. By leveraging local data processing capabilities, these methods reduce dependence on cloud resources and enhance data privacy \cite{shao_2020}. 

A recently prominent approach to distributed learning is split learning (SL), recently considered in \cite{gupta_2018, vep_2018, poirot_2019, lin_2024}. SL emerged as an effective method that allows edge devices to delegate training workloads to edge servers while keeping raw data locally stored \cite{lin_2024}. SL employs layer--wise splitting, wherein a deep neural network is fragmented into smaller networks, distributing network architecture and respective parameters across multiple nodes \cite{tuli_2022}.  In the context of IoT networks, the neural network architecture is usually partitioned into two parts, with one part deployed at the IoT edge device and the other at the edge server. Each edge device performs forward computation on its designated portion of the neural network, transmitting computed intermediate results to the succeeding edge device or a server for additional computation and estimation \cite{yang_2023}. In this manner, the tasks that require substantial computational resources are shifted from IoT edge devices to nearby edge servers. This approach aims to minimize latency in processing and mitigate the risk of data exposure, contrasting with the exclusive reliance on cloud computing \cite{itahara_2022}.

The practical realisation of SL strategy in a real--world IoT system faces significant challenges due to unpredictable wireless channel conditions. Wireless signal propagation effects result in stochastic channel fluctuations that affect data transmission between edge devices or between an edge device and an edge server \cite{cao_2023}. As a result, intermediate messages (representations) exchanged between the devices may be affected by channel-induced impairments, leading to potentially poor quality inference. While this issue has received considerable attention in recent literature leading numerous authors to propose various solutions \cite{itahara_2022, yang_2023, shao_2020_1,  krouka_2021, samikwa_2022, baccareli_2021, ares_2022, wu_2023, tuli_2022}, majority of these papers have focused on specific aspects (e.g., channel models) of the system. To the best of our knowledge, a broader, more comprehensive analysis of the SL system behavior in the context of communication--aware SL is still missing. Moreover, the proposed design targets a wide range of IoT systems operating with time series data, addressing the most common practical IoT scenario that is not explored in the literature.

\begin{table*}[tbhp]
\caption{Communication-aware SL in IoT Networks - SotA.}
\begin{center}
\begin{tabular}{|c|c|c|c|c|c|c|c|}
\hline
Paper & Comm.--aware SL approach &Erasure channel & AWGN channel & Early--exit & Heterogeneous setup & Time series &Compression\\
\hline
\cite{gupta_2018, vep_2018, poirot_2019}&&& & &\checkmark & &\checkmark\\
\hline
\cite{itahara_2022}&\checkmark&\checkmark& & & & &\checkmark\\
\hline
\cite{shao_2020_1,  krouka_2021}&\checkmark& &\checkmark & & & &\\
\hline
\cite{shao_2020}&\checkmark& &\checkmark & & & &\checkmark\\
\hline
\cite{yankowski_2023}&\checkmark& &\checkmark &\checkmark & & &\checkmark\\
\hline
\cite{samikwa_2022, baccareli_2021}&\checkmark& & &\checkmark & & &\\
\hline
\cite{yang_2023}&\checkmark& &\checkmark &&\checkmark & &\\
\hline
\cite{disnet_2024, ares_2022, wu_2023, tuli_2022}&\checkmark& & & &\checkmark &&\checkmark\\
\hline
\cite{koda_2020}&\checkmark& & & && \checkmark&\checkmark\\
\hline
\cite{jiang_2022}&& & & &\checkmark& \checkmark&\checkmark\\
\hline
COMSPLIT&\checkmark&\checkmark & \checkmark& \checkmark&\checkmark& \checkmark&\checkmark\\
\hline
\end{tabular}
\label{table_sota}
\end{center}
\end{table*}

In this paper, we propose a novel communication-aware SL design named COMSPLIT which is tailored to generic scenarios characterised by heterogeneous IoT platforms. By incorporating an additional layer to emulate channel conditions during the SL training phase, our approach demonstrates adaptability to various channel environments. This flexibility proves particularly advantageous in systems characterized by a mix of edge devices with diverse computational capabilities. Such networks exhibit a potential for processing data from diverse source modalities, enabling them to prioritize certain data sources according to their importance for specific tasks. For instance, in surveillance systems, video data may be deemed significantly more important than certain sensor measurements. Furthermore, the integration of an early--exit strategy \cite{szegedy_2015, surat_2017} enhances system autonomy, reducing latency and communication overhead \cite{yankowski_2023, baccareli_2021}, and allowing for the determination of an optimal transmission strategy based on predefined performance criteria. We demonstrate that the proposed framework significantly outperforms vanilla SL approaches (trained under ideal channel assumption), but also define different trade-offs for server, which may adapt its operations dynamically according to real--time constraints and network conditions, thereby optimizing resource utilization and overall system efficiency. The paper contributions can be summarized as follows:
\begin{itemize}
    \item We introduce a novel communication-aware SL solution that outperforms vanilla SL algorithms under identical testing conditions. Unlike existing approaches, COMSPLIT addresses a wider context by demonstrating robustness across different channel models. An additional level of adaptability to varying channel conditions is obtained via utilization of an early-exit strategy.  
    \item The COMSPLIT approach is extended to a heterogeneous IoT environment, accommodating edge devices with varying computational capabilities. Consequently, it can be applied in multi-modal scenarios, offering reduced processing overhead as the same server neural network is utilized across edge devices of heterogeneous performance.
    \item Integration of time series data and recurrent neural networks (RNNs) with long short-term memory (LSTM) within the COMSPLIT architecture goes beyond existing approaches, addressing diverse scenarios closely linked to IoT deployment in real--world monitoring systems.
    \item In this work, we expand the conventional additive white Gaussian noise (AWGN) channel to two-state AWGN model that includes occasional deep fades that may affect random symbols within the transmitted representation. Under such channel behavior, COMSPLIT demonstrates graceful robustness and the capability to adapt to channel behaviour, ensuring reliable performance in challenging wireless environments.
    \item As a real--world demonstration, the COMSPLIT design is applied to a practical scenario of water quality monitoring using real--world data collected from the Danube river near the city of Novi Sad (Serbia). The performance of the IoT system, based on the core principles of COMSPLIT, remains consistent with that observed in the generic case. This indicates that the proposed design is well--suited to address challenges in real--world IoT systems and holds significant promise for future deployment in various IoT applications.
\end{itemize}

\subsection{Related Work}
\label{sec:related}
Distributed learning and inference algorithms in IoT networks have garnered significant attention in recent years due to their numerous advantages over centralized approaches. These algorithms alleviate the workload on central processors, preserve user data privacy and reduce overall system latency\cite{cao_2023}. While federated learning (FL) \cite{kon_2016} has been considered robust solution, the increasingly high demands in terms of computational capabilities and communication costs \cite{kairouz_2021} have hindered its deployment at IoT network edge devices \cite{khan_2021, chen_2023}.

The split learning (SL) paradigm has recently emerged as an innovative distributed learning framework, enhancing privacy and targeting resource-constrained devices\cite{gupta_2018, vep_2018, poirot_2019, lin_2024}. Many practical questions arise regarding the optimal neural network splitting strategy \cite{lin_2024}, model placement \cite{yan_2022}, and additional privacy protection \cite{pham_2023}. In addition, authors in \cite{cao_2023} underline the pressing demand for creation of communication-aware distributed learning algorithms. These algorithms should effectively minimize communication costs while maintaining satisfactory learning and optimization performance. Next, we provide a comprehensive overview of communication-aware SL methods in the recent literature, with the summary presented in Table \ref{table_sota}. 

In \cite{itahara_2022}, the proposed approach enhances communication-oriented distributed inference in IoT by fine-tuning SL models to handle packet loss, thus ensuring accurate predictions with low latency in lossy networks. More precisely, packet erasure channel effects are emulated by the introduction of an additional dropout layer during the SL training phase. AWGN channel is incorporated into SL model through additional noise layer in \cite{shao_2020_1, shao_2020, krouka_2021, yankowski_2023}. In \cite{shao_2020_1}, a three-step framework is introduced for efficient inference involving selecting the on-device model, compressing it for reduced computation and communication overhead, and encoding intermediate features for communication reduction. Additionally, \cite{shao_2020} proposes BottleNet++, an end-to-end architecture featuring an encoder, a non-trainable channel layer, and a decoder. BottleNet++ leverages lightweight convolutional neural networks (CNNs) to implement joint source-channel coding while accounting for channel noise, thereby enhancing feature compression and transmission efficiency. In \cite{krouka_2021}, the authors present analog implementations of split learning optimized for noisy wireless channels and constrained transmission power. 

Distributed inference in IoT systems usually introduces significant communication overhead due to the transmission of neural network parameters between devices, consequently leading to significant latency \cite{shi_2022}. In order to overcome this drawback, several papers proposed early--exit strategy in IoT networks \cite{yankowski_2023, samikwa_2022}. This strategy, originally presented in \cite{szegedy_2015, surat_2017}, incorporates additional outputs into the neural network architecture, enabling local estimation at edge devices. More precisely, communication savings is investigated in \cite{yankowski_2023} where  an early--exit mechanism is introduced for collaborative inference systems deployed at the wireless edge. The core of the proposed approach is the transmission--decision mechanism, which utilizes channel state information to make binary decisions about whether to accept early--exit output or transfer data to the edge server for further processing. An adaptive, energy-efficient inference scheme for IoT networks, which dynamically allocates computation between IoT devices and edge servers, aiming to minimize prediction latency and energy consumption is presented in \cite{samikwa_2022}.

Behaviour of distributed inference in IoT networks comprised of different edge devices with heterogeneous computational capabilities is examined in \cite{disnet_2024, ares_2022, wu_2023}. The strategy detailed in \cite{disnet_2024} for distributed micro-split deep learning in heterogeneous IoT systems enhances inference speed and decreases energy consumption by combining vertical and horizontal DNN partitioning, enabling flexible, distributed, and parallel execution of neural network models across various IoT devices. It takes computing and communication resources of IoT devices and network conditions into account to facilitate resource-aware cooperative DNN inference. The same authors introduce a scheme for efficient model training in IoT systems \cite{ares_2022}, which accelerates training in resource-constrained devices, minimizes the impact of stragglers, and accounts for time-varying network throughput and computing resources. Further reduction in training latency is achieved in \cite{wu_2023}, where devices are partitioned into multiple clusters for parallel training of device-side models within each cluster, followed by aggregation and sequential training of the complete AI model across clusters. 

Finally, the deployment and maintenance of distributed (split) learning/inference algorithms in IoT environments pose several practical challenges. Beyond the well-known issue of communication overhead \cite{shi_2022, gao_2021}, significant concerns stem from traditional resource management issues such as bandwidth and power. These resources must be efficiently managed to meet performance requirements, including bit error rate, energy and spectral efficiencies, outage probability, and network throughput \cite{le_2024}. From a real-time processing perspective, \cite{bian_2022} highlights the daunting scheduling problems posed by modern hardware, such as heterogeneous and multiprocessor platforms. They underscore the pressing need for real-time capability and scalability, especially in resource-constrained environments, to ensure seamless and efficient operation of IoT systems. The challenge is further compounded by the fact that the failure of a single machine can halt the entire training process. While this may be manageable with fewer nodes, as the number of nodes increases, the likelihood of any node becoming unavailable grows, leading to frequent and near-continuous interruptions in the training process \cite{verb_2021}.

The structure of the paper is organized as follows. Section II  provides an in-depth introduction to split learning and inference, highlighting the application of LSTM in this context. In Section III, we describe the system model, focusing on the various scenarios that are key to this work. Section IV presents the proposed design framework, detailing the implementation aspects. The performance evaluation, along with experimental setup and training procedure, is thoroughly examined in Section V. Finally, Section VI concludes the paper, summarizing the main findings and future work.

\section{Background}

\subsection{Split Learning \&\ Inference}
\label{SL}

\begin{figure}
	\centering
	\includegraphics[width=1\linewidth]{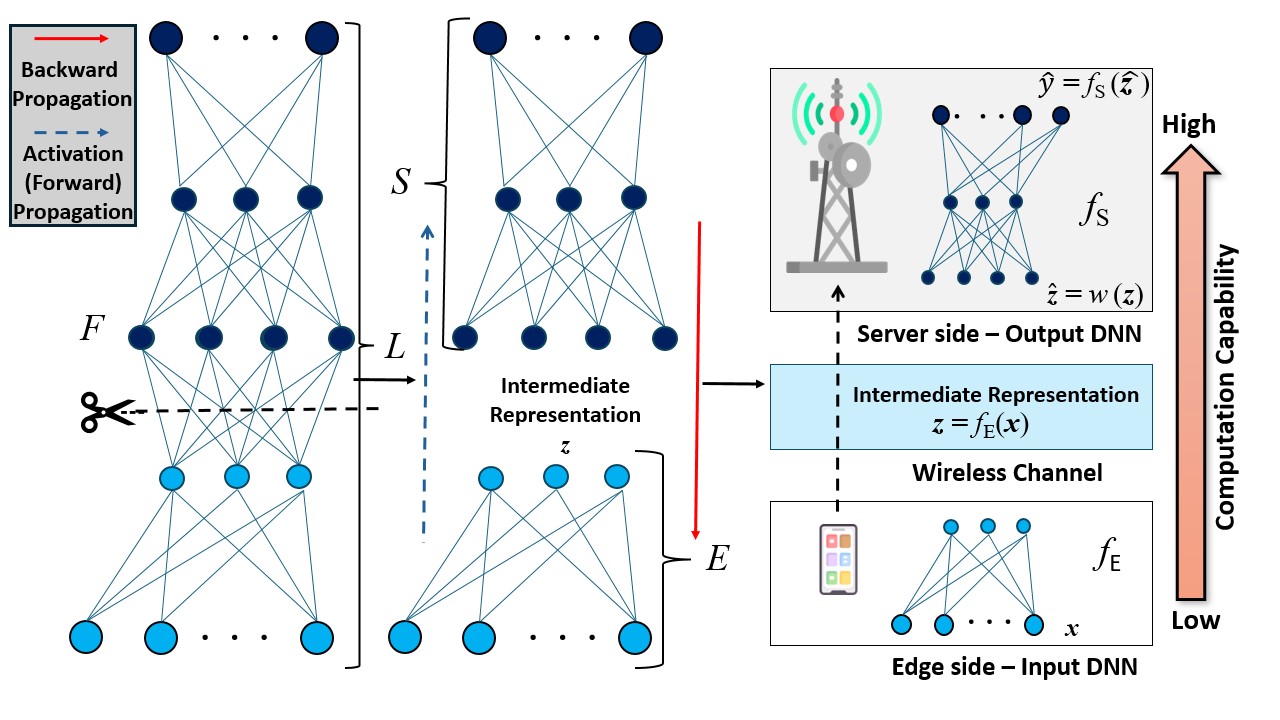}
 \vspace{-5mm}
	\caption[figure caption]{Split learning/inference architecture with edge device and server sub--networks implementation.\footnotemark[2]}
	\label{fig_split}
\end{figure}

Split learning (SL) \cite{gupta_2018, vep_2018, poirot_2019, lin_2024} facilitates distributed learning by partitioning a neural network $F$, consisting of $L$ layers, into sequential layers spread across multiple participants. An SL example, consisting of an edge device and a server, is shown in Fig. \ref{fig_split}. In this paradigm, the edge device collaborates with the server to train a shared deep neural network by securely aggregating its training dataset, while the server orchestrates and guides the network's training process assuming the bulk of computational responsibilities. As a consequence, different sub--networks of $F$ implemented at the edge device and the server consist of $E$ and $S$ layers, respectively, where $E<S$ and $L=E+S$. This distributed approach not only accelerates convergence by enabling processing of different sub--networks across multiple nodes but also mitigates bandwidth constraints associated with transferring original raw data \cite{vep_2018}. 

The central idea behind split learning involves segregating the processes of offline model training and online inference (also referred to as split inference). During sequential training, which is performed through vertically--distributed back propagation \cite{langer_2020}, data remains localized within individual edge devices to prevent the transmission of raw information $\boldsymbol{x} \in \mathbb{R}^{N}$ across the network. The sub--networks at the edge device and the server, which are jointly optimized, can be defined by functions  $f_{\text{E}}:\mathbb{R}^{N}\rightarrow\mathbb{R}^{M}$ and  $f_{\text{S}}:\mathbb{R}^{M}\rightarrow\mathbb{R}$, where $N$ and $M$ represent dimensions of raw data and intermediate representation $\boldsymbol{z}=f_{\text{E}}(\boldsymbol{x})$, ($\boldsymbol{z}\in \mathbb{R}^{M}$), respectively, and $M<N$\footnotemark[1] \footnotetext[1]{As we focus on regression problems, we have defined $f_{\text{S}}:\mathbb{R}^{M}\rightarrow\mathbb{R}$ (with only one output value). Furthermore,  $\boldsymbol{z}$ can be understood as a compressed representation of $\boldsymbol{x}$. These assumptions
 may not hold in general neural network architecture.}. Neural network $F$ is consequently represented as $F=(f_{\text{E}}, f_{\text{S}})$. During the activation (forward propagation), the edge device sub--network output $\boldsymbol{z}=f_{\text{E}}(\boldsymbol{x})$ is sent to the server (Fig. \ref{fig_split} blue arrow). At the server side, prediction based on the received intermediate representation $\hat{\boldsymbol{z}}$ is obtained as  $\hat{y}=f_{\text{S}}(\hat{\boldsymbol{z}})$. Subsequently, the appropriate loss function, such as categorical cross-entropy for classification tasks or mean squared error for regression tasks, is calculated \cite{pasquini_2021}. Such training facilitates collaborative learning without compromising data privacy, as model updates encoded as gradients are iteratively exchanged during the backward pass between the server and the edge device (Fig. \ref{fig_split}, red arrow). In the case of supervised learning, as explored in this work, the server has access to labels, although this requirement can be circumvented by updating the loss function at the edge side, as proposed by \cite{vep_2018, pasquini_2021}. 

\footnotetext[2]{In our work, $\hat{y} \in \mathbb{R}$, thus network architecture has only one neuron in the output layer. In this figure we presented general case.}

In the online split inference phase, split learning further refines its efficiency by leveraging pre--trained models deployed across multiple devices. Input data undergoes initial processing locally at the edge device, generating intermediate representation $\boldsymbol{z}$, which is transmitted to a centralized server for aggregation and final inference \cite{vep_2018}. In our setup ($M<N$), the intermediate representation can be understood as a compressed information extracted from the input data, facilitating faster and more efficient inference process \cite{zhao_2024, itahara_2022, shao_2020}.

By decentralizing the training/inference process, split learning/inference minimizes bandwidth consumption, compared to other distributed approaches\cite{singh_2019}. Moreover, the computational load on edge devices is lighter than that induced by federated learning \cite{kairouz_2021}, as they only need to perform forward/backward propagation on a smaller portion of the network rather than on the entire structure. 
 Despite recent concerns about potential vulnerabilities regarding privacy  \cite{pasquini_2021}\cite{abua_2020},  split learning and its extensions \cite{thapa_2022}\cite{jeon_2020} offer a robust framework for enhancing privacy in sensitive applications, contributing to the design of privacy--preserving machine learning methods \cite{gupta_2018}.

\subsection{Split Learning--Based LSTM}

LSTM neural networks represent an enhanced version of RNNs and serve as a cornerstone algorithm for learning on time-series data \cite{greff_2017}. The initial attempts of integration of the split learning/inference paradigm into LSTM neural networks posed different challenges, leading researchers to explore alternative approaches. Multiple works have reported the utilization of 1D-CNN as a replacement for LSTMs in addressing these complexities and achieving effective model implementation \cite{abua_2020, Gao_2020, zhang_2024}. Recent research presents novel approaches that efficiently integrate split learning into LSTM networks \cite{koda_2020, jiang_2022, abedi_2023}. These methods directly embed the split learning paradigm within LSTM architectures, offering innovative strategies to overcome implementation hurdles. 

\begin{figure}
	\centering
	\includegraphics[width=1\linewidth]{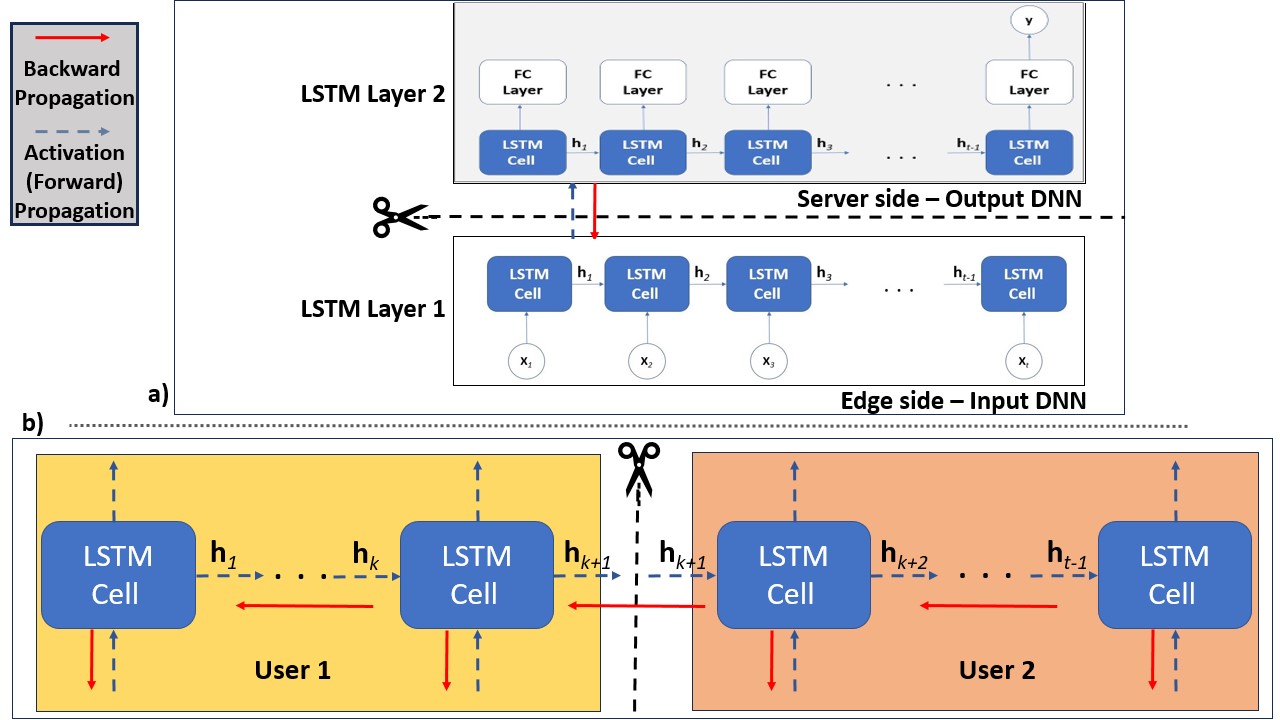}
\vspace{-5mm}	
 \caption{LSTM--based split learning/inference algorithms: a) LSTMSPLIT \cite{jiang_2022}; b) Fedsl \cite{abedi_2023}.}
	\label{LSTM_split}
\end{figure}

In the first approach \cite{jiang_2022}, whose main principles we reuse in this paper, the authors introduced \textit{LSTMSPLIT} algorithm, which vertically splits the LSTM neural network, as depicted in Fig. \ref{LSTM_split}a (LSTM hidden states are represented as $\textbf{h}$). Such a network should have at least two layers and the entire input sequence is stored at the edge device side. The remaining steps follow the procedure presented in Section \ref{SL}, where the intermediate representation $\boldsymbol{z}$ is transmitted from the LSTM layer at the edge device side to the LSTM layer at the server side, while update gradients are transmitted in the opposite direction (Fig. \ref{LSTM_split}, blue and red arrows, respectively).

The second approach utilizes the sequential (or recurrent) structure of the LSTM network presented as a directed graph over a temporal sequence \cite{abedi_2023}. Specifically, one LSTM layer is distributed over multiple edge devices, as shown in Fig.~\ref{LSTM_split}b. This entails dividing it into sub--networks, each accessible and trained on an individual edge device, to handle one segment of a multi-segment training sequences, as presented in Fig. \ref{LSTM_split}b for two users.  In addition to edge device-server communication, multiple edge devices also communicate and share parameters of sub-networks to facilitate appropriate inference, which naturally leads to interconnection between the split learning and the federated learning paradigm (which falls outside the scope of this work). 

\section{System Model}
\label{model}
We address a collaborative inference problem tailored to a standard IoT network setup handling time series input data. The IoT network comprises an edge device and a server, each with distinct computational capabilities, as illustrated in Fig.~\ref{fig_split} (rightmost part). The workflow assumes that the edge device gathers raw data $\boldsymbol{x} \in\mathbb{R}^{N}$ from various sensors or cameras and processes it into a compressed intermediate representation before transmitting it to the server for generating predictions. From the split learning/inference perspective, akin to the concepts detailed in Section \ref{SL}, a single neural network $F$ is partitioned into two sub-networks ($f_{\text{E}}$ and $f_{\text{S}}$) with differing complexities ($\mathcal{C}(f_{\text{E}})<\mathcal{C}(f_{\text{S}})$), deployed at the edge device and the server, respectively (Fig. \ref{fig_split}).

The edge device holds a local dataset $\mathcal{D}$ consisting of $P$ individual instances defined as $(\boldsymbol{x}, y)$, where $\boldsymbol{x} \in \mathbb{R}^N$ denotes the raw data point and $y \in \mathbb{R}$ is the corresponding label. During the forward propagation phase, the raw data is fed into the edge device sub-network, which generates a compressed intermediate representation $\boldsymbol{z}=f_{\text{E}}(\boldsymbol{x})$ comprising $M$ real-valued symbols ($\boldsymbol{z}\in\mathbb{R}^M$, see Sec. \ref{SL}). These symbols are transmitted through a wireless channel $\mathcal{W}$, resulting in the channel output $\hat{\boldsymbol{z}}=\mathcal{W}(\boldsymbol{z}), \hat{\boldsymbol{z}}\in\mathbb{R}^M$, that is available at the server input. The prediction $\hat{y}=f_{ \text{S}}(\hat{\boldsymbol{z}}), \hat{y} \in \mathbb{R}$, is obtained as an output at the server side (Fig. 1, rightmost part). Depending on the defined scenario, an appropriate loss function is computed at the server side, such as mean squared error for regression tasks:

\begin{align}
\label{loss}
    \mathcal{L}(y, \hat{y})=\frac{1}{P}\sum_{\mathcal{D}}(y-\hat{y})^2
\end{align}
During the backward propagation phase, gradients are calculated based on the loss function, and they propagate from the server towards the edge device, following the reverse direction of the neural network (red arrows in Fig. \ref{fig_split}). This way, both sub-networks $f_{\text{E}}$ and $f_{\text{S}}$ undergo a joint optimization and updating process. The joint optimization process typically involves updating the parameters of both sub-networks using an optimization algorithm such as stochastic gradient descent (SGD) or its variants (Adam) \cite{adam}.

\begin{figure}
	\centering
	\includegraphics[width=1\linewidth]{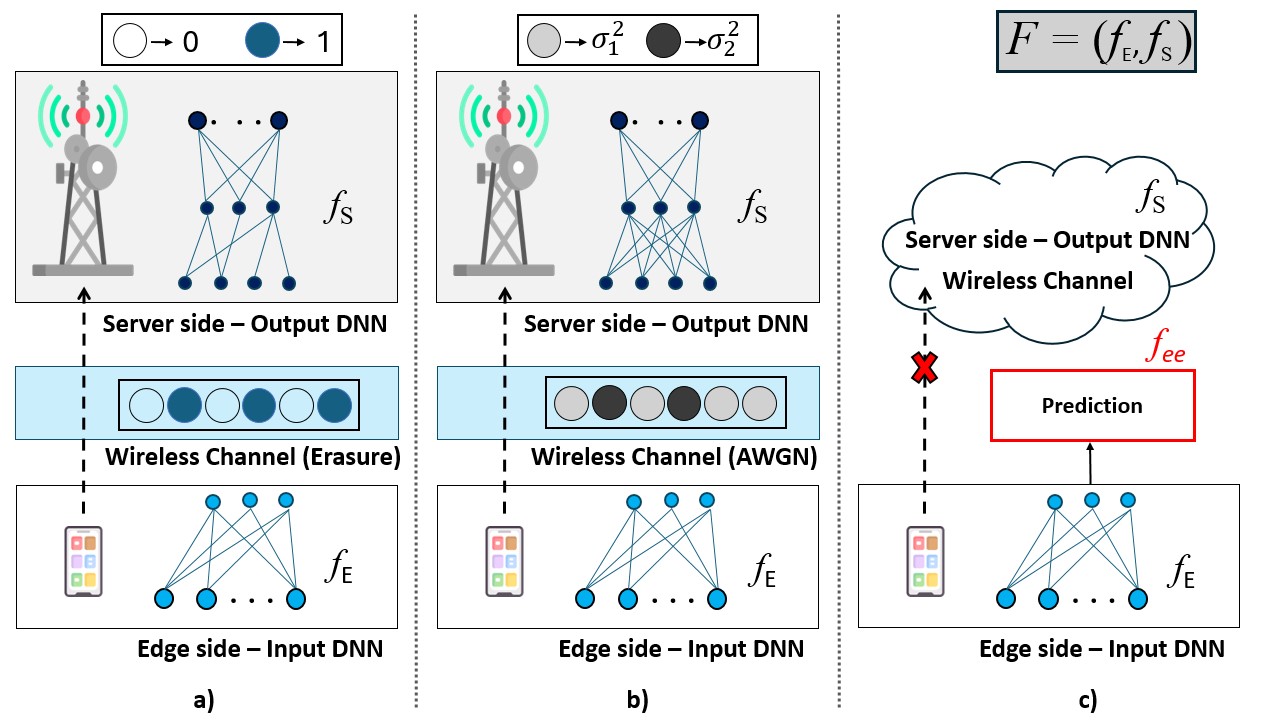
 }
 \vspace{-4mm}
	\caption{Communication-aware split learning/inference model incorporating: a) Erasure channel; b) AWGN channel; c) Early exit \cite{yankowski_2023}.}
	\label{fig_scen}
\end{figure}

The main practical challenge of the above--described system is represented by the random nature of the wireless channel (Fig. \ref{fig_split}), which can significantly damage the intermediate representation $\boldsymbol{z}$ and deteriorate the overall system performance \cite{yang_2023, itahara_2022, yankowski_2023}. Our goal here is to enable communication--efficient split learning/inference algorithm which, for a given channel $\mathcal{W}$, optimizes the mean squared error (MSE) defined as $\mathcal{E}=1/|\mathcal{D}|\sum_{\mathcal{D}}(y-\hat{y})^2$, where $|\cdot|$ represents cardinality of a set. To comprehensively analyze the behavior of the LSTM-based split learning/inference model in IoT networks under various channel conditions, we define scenarios depicted in Figs. \ref{fig_scen}a, \ref{fig_scen}b, \ref{fig_scen}c and \ref{fig_hIoT}.

\subsection{Split Inference Over Erasure Channel}
\label{errasure_chan}
In the first scenario, we model the wireless channel between the edge device and the server as a (symbol) erasure channel. More precisely, individual symbols from the intermediate representation $\boldsymbol{z}$ are either erased with probability $p$ (empty circles in Fig. \ref{fig_scen}a) or arrive correctly with probability $1-p$ (blue circles in Fig. \ref{fig_scen}a). We represent an erasure channel as an $M$--dimensional binary vector $\boldsymbol{q} \in\{0,1\}^{M}$ whose entries are sampled from a Bernoulli distribution with a probability of one equal to $1-p$ (we note that this is a symbol-level adaptation of a similar model considered in \cite{itahara_2022}). The received intermediate representation can be defined as $\hat{\boldsymbol{z}}=\mathcal{W}(\boldsymbol{z})=\boldsymbol{z}\odot \boldsymbol{q}$, where $\odot$ denotes element--wise multiplication. We note that using erasure channel as a model for IoT communications is not uncommon \cite{quin_2024, julio_2020}, as it usually models deep fading or strong interference affecting particular symbols. In the following scenario, we refine the erasure channel model with an appropriate variant of a Gaussian channel.  

\subsection{Split Inference Over AWGN Channel}
\label{AWGN}
The second scenario involves the utilization of an additive white Gaussian channel (AWGN) as the wireless channel $\mathcal{W}$ between the edge device and the server (see Fig. \ref{fig_scen}b). We represent the AWGN channel as an $M$--dimensional real vector $\boldsymbol{n} \in\mathbb{R}^{M}$. The resulting server input is defined as $\hat{\boldsymbol{z}}=\mathcal{W}(\boldsymbol{z})=\boldsymbol{z}+\boldsymbol{n}$. Note that we slightly extend the conventional AWGN channel (used in recent split learning studies \cite{yankowski_2023, lan_2021}) by introducing occasional deep fades that more intensively affect certain randomly selected $\boldsymbol{z}$ symbols. Specifically, $\boldsymbol{n}$ contains $M_1$ independent and identically distributed (i.i.d.) samples of a Gaussian random variable $\mathcal{N}_1(0,\sigma_1^2)$ and $M_2$ i.i.d. samples of a Gaussian random variable $\mathcal{N}_2(0,\sigma_2^2)$ (light and dark gray circles in Fig. \ref{fig_scen}b), where $\sigma_1^2<\sigma_2^2$ and $M=M_1+M_2$. Furthermore, the positions of symbols sampled from two different distributions within vector $\boldsymbol{n}$ are selected randomly (among all possible such sequences), as illustrated in Fig. \ref{fig_scen}b. This scenario is of particular interest in IoT systems, where deep fades can result from intermittent obstacles arising between the edge device and server. For example, these deep fades can reflect occasional movement of a unmanned aerial vehicle (UAV) when it is used for coverage enhancement \cite{zhang_2019}.

\subsection{Split Inference with Early--Exit}
\label{sec:early}
In IoT networks, edge devices often operate in challenging channel conditions, such as those in rural environmental or underground mining scenarios. This naturally raises the question: What is the impact on split inference performance if the intermediate representation experiences adverse channel conditions? If the channel is in a bad state, how useful is to deliver (highly corrupted) intermediate representation to the server? Moreover, the possible trade-off between inference reliability and inference delay can be significant for critical applications \cite{baccareli_2021}, such as those in which decisions must be made with minimal latency (emergency response systems, fires or earthquakes).

To address the aforementioned problems, we consider implementing an early-exit strategy applied at edge devices \cite{szegedy_2015, surat_2017}. The main idea is adding an additional neural network output to the edge device (Fig. \ref{fig_scen}c), denoted as $f_{\text{ee}}$, that provides earlier prediction output without requiring full neural network processing. For this approach, we assume that the edge device knows either exactly or approximately the channel state information, based on which it decides whether to send intermediate representation $\boldsymbol{z}$ to the server or perform inference locally, i.e., $\hat{y}_{ee}=f_{ee}(\boldsymbol{z})$. The overall loss function (Eq. (\ref{loss})) is revised in order to include the early-exit scenario:
\begin{align}
\label{loss_early}
    \mathcal{L}_{ee}=\mathcal{L}(\hat{y}, y)+\mathcal{L}(\hat{y}_{ee}, y)
\end{align}

Several recent papers attempted to implement this strategy at the edge device, employing various transmission strategies \cite{yankowski_2023, baccareli_2021, samikwa_2022} (Section \ref{sec:related}). The consensus from these studies is that this approach reduces communication overhead, latency, and energy consumption, albeit with a performance degradation compared to the case without early exit.

\subsection{Split Inference with Heterogeneous IoT Devices}
\label{hetIoT}
\begin{figure}[t]
	\centering
	\includegraphics[width=0.98\linewidth]{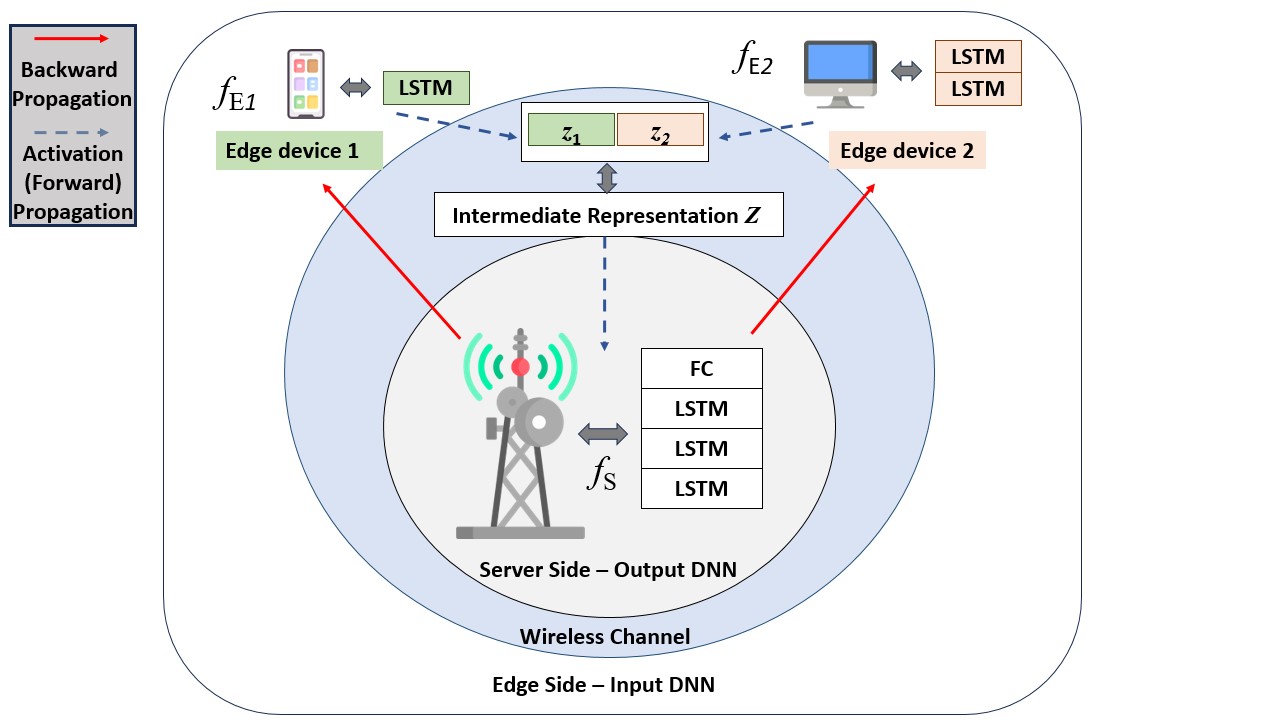}
 \vspace{-2mm}
	\caption{Split learning/inference model utilizing multiple edge devices with varying computational capabilities.}
	\label{fig_hIoT}
\end{figure}

In the preceding scenarios, the communication system contains a single edge device and a server (Fig. \ref{fig_scen}a, b, and c). Building upon this, we expand the analysis of split learning/inference behavior in IoT networks by integrating multiple edge devices that communicate simultaneously with the server, as shown in Fig. \ref{fig_hIoT}. To accommodate additional devices,  we slightly revise the system model presented in Section \ref{model}.

%Each device produces an intermediate representation $\boldsymbol{z}_i=f_{\text{E}_i}(\boldsymbol{x}_i)$ , which is sent through the same wireless channel $\mathcal{W}$ (e.g., using time-division scheduled access) to the server. At the server, the individual edge device intermediate representations are concatenated as $\boldsymbol{Z}=(\boldsymbol{z}_1, \boldsymbol{z}_2,\ldots,\boldsymbol{z}_C)$ ($\boldsymbol{Z}\in\mathbb{R}^{C\times M}$),  as depicted in Fig. 4 for two devices

We assume a conventional scheduled access strategy in which an IoT network consists of $C$ edge devices. Each device generates a local dataset $\mathcal{D}_i$, where $i\in{1,2,\ldots,C}$, and each dataset consists of  $(\boldsymbol{x}_i, y_i)$ instances. %At the output of each device, an intermediate representation $\boldsymbol{z}_i=f_{\text{E}_i}(\boldsymbol{x}_i)$ is produced. %The concatenation of all individual edge device intermediate representations $\boldsymbol{Z}=(\boldsymbol{z}_1, \boldsymbol{z}_2,\ldots,\boldsymbol{z}_C)$ ($\boldsymbol{Z}\in\mathbb{R}^{C\times M}$) is sent through the same wireless channel $\mathcal{W}$ towards the server (e.g., using time-division scheduled access), as depicted in Fig. \ref{fig_hIoT} for two devices. and sent to the server using time-division scheduled access
 An intermediate representation $\boldsymbol{z}_i=f_{\text{E}_i}(\boldsymbol{x}_i)$ is generated by each device and sent to the server using time-division scheduled access. There, the individual edge device intermediate representations are concatenated as $\boldsymbol{Z}=(\boldsymbol{z}_1, \boldsymbol{z}_2,\ldots,\boldsymbol{z}_C)$ ($\boldsymbol{Z}\in\mathbb{R}^{C\times M}$),  as depicted in Fig. \ref{fig_hIoT} for two devices. For each device-to-server transmission, we consider channel model defined in Section \ref{errasure_chan}. At the server side, predictions are derived for each of the $C$ edge devices using $\hat{\boldsymbol{Z}}=\mathcal{W}(\boldsymbol{Z})=(\hat{\boldsymbol{z}}_1, \hat{\boldsymbol{z}}_2, \ldots, \hat{\boldsymbol{z}}_C)$ as an input, i.e., $(\hat{y}_1, \hat{y}_2, \ldots, \hat{y}_C)=f_{\text{S}}(\hat{\boldsymbol{z}}_1, \hat{\boldsymbol{z}}_2, \ldots, \hat{\boldsymbol{z}}_C)$. The loss function is modified as follows:
\begin{align}
\label{loss_het}
    \mathcal{L}_{het}=\sum_{i=1}^{i=C}\mathcal{L}_{i}(y_i, \hat{y}_i)
\end{align}
As a consequence, $C$ different edge sub-networks are jointly optimized with one server sub-network, and the overall scenario can be defined with $(\{f_{\text{E}_i}\}_{i=1,\ldots,C}, f_{\text{S}})$ pairs. The primary objective is to optimize the overall error profile of the system, denoted as $\mathcal{E}_{\text{sys}} = \{\mathcal{E}_1, \mathcal{E}_2, \ldots, \mathcal{E}_C\}$, where $\mathcal{E}_i$ is average MSE error, defined in Section \ref{model}. 

This scenario not only brings us one step closer to real-world IoT networks but also offers several advantages. In particular, by introducing various edge devices, we establish a flexible setup for investigating multi-modal learning, where each device can process data from different source modalities (as discussed in \cite{koda_2020} and \cite{jiang_2022}). Additionally, we assume heterogeneous computational capabilities across different edge devices (Fig. \ref{fig_hIoT}), i.e., for any devices $i$ and $j$, $\mathcal{C}(f_{\text{E}_i}) \neq \mathcal{C}(f_{\text{E}_j})$. Due to uneven computing capabilities of individual devices, edge sub-network complexity at different devices could be adjusted, e.g., by placement of different number of neural network layers. %The neural sub--network implemented on the server side manages the importance of data acquired from various sources (edge devices) during the training/inference phase.

\section{COMSPLIT: Communication--Aware Split Inference Techniques}

In this section, a communication-aware LSTM-based split learning/inference design framework, entitled COMSPLIT, is proposed. In COMSPLIT, one or more edge devices communicate with the server, transmitting intermediate representations over different wireless channels, thereby introducing distortions and erasures into the transmitted data. The primary objective of this framework is to integrate diverse channel conditions during the offline split learning phase, thereby learning an optimal solution for constructing intermediate representations and enabling the design of a robust server sub--network capable of mitigating channel perturbations during the online split inference phase.   

\subsection{Baseline COMSPLIT Model}
\label{COMSPLIT}
To address the convergence of the neural network learning process with various communication effects that may emerge during transmission, we choose to incorporate an additional layer into the neural network model $F=(f_{\text{E}}, f_{\text{S}})$ (Section \ref{SL}), as illustrated in Figs. \ref{fig_scen}a and \ref{fig_scen}b. This layer emulates channel effects and modifies specific layer outputs (intermediate representation) during activation (forward) propagation in the offline network training phase. As a result, the server sub--network will be trained to handle distorted intermediate representations received during the online inference phase. Here, we analyse integration of two channel models, presented in the Sections \ref{errasure_chan} and \ref{AWGN}.   

The behavior of a symbol-based erasure channel (Section \ref{errasure_chan}), as described in \cite{itahara_2022}, can be replicated by introducing a conventional dropout layer between the edge and server sub-networks, as depicted in Fig. \ref{fig_scen}a. Dropout is a well-known regularization technique utilized to prevent neural networks from overfitting \cite{hinton_2012}. It randomly drops hidden units with a specified probability $p$ (empty circles in Fig. \ref{fig_scen}a), while the remaining units remain unchanged (blue circles in Fig. \ref{fig_scen}a). From the communication perspective, if we represent the dropout layer with a binary vector $\boldsymbol{q} \in \{0,1\}^M$ and suppose that hidden units can be understood as a symbols of intermediate representation, the dropout layer becomes a convenient model for erasure channel.    

On the other hand, the AWGN channel is implemented as a non-trainable noise layer, as described in \cite{OShea_2017}, which adds Gaussian noise independently to each symbol in the intermediate representation (Fig. \ref{fig_scen}b, light and dark gray circles). As emphasized in Section \ref{AWGN}, occasional deep fades are considered, meaning that this layer comprises symbols sampled randomly from two Gaussian distributions with zero mean and variances $\sigma_1^2$ (light gray circles) and $\sigma_2^2$ (dark gray circles). To calculate signal--to--noise (SNR) ratio, we reuse and adjust the formula from \cite{yankowski_2023} and \cite{OShea_2017} to fit the scenario of interest. Specifically, if we define intermediate representation as an M--value vector $\boldsymbol{z}=\{z_1, z_2, \ldots, z_M\}$, SNR is calculated as $SNR=R/\sigma^2$, where $R=1/M\sum_{k=1}^{k=M}z_k^2<1$. Moreover, it is important to note that the random nature of the channel can be seen as a form of regularization because the receiver never encounters the same training example twice. Consequently, it is highly unlikely for the neural network to overfit \cite{Dorner_2018}. 

\subsection{COMSPLIT Model with Early--Exit}
\label{early_weight}
The above scenario is further extended to incorporate an early--exit strategy. Specifically, an additional fully--connected (FC) layer is introduced at the output of the edge device, as illustrated in Fig. \ref{fig_scen}c (red square, $f_{ee}$). This layer facilitates the generation of less accurate predictions as compared to the server side prediction however without the need for transmitting intermediate representations to the server side. The literature presents various schemes concerning the decision--making process at the edge device regarding either transmitting data or making early predictions. For instance, in \cite{yankowski_2023}, channel conditions are estimated by neural network and decision is made in order to minimize communication overhead, whereas in \cite{samikwa_2022}, the emphasis is on optimizing overall IoT network latency and energy consumption. 

During the offline split learning phase, collected raw data $\boldsymbol{x}$ is first fed into the edge device sub--network in order to produce an intermediate representation $\boldsymbol{z}=f_{\text{E}}(\boldsymbol{x})$. The intermediate representation is simultaneously transmitted via  wireless channel towards the server ($\hat{y}=f_{\text{S}}(\mathcal{W}(\boldsymbol{z}))=f_{\text{S}}(\hat{\boldsymbol{z}}$)), and passed locally through an additional FC layer at the edge device ($\hat{y}_{ee}=f_{ee}(\boldsymbol{z})$). The overall loss function is the sum of loss calculated at the server and at the edge device (see Eq. (\ref{loss_early})). During backpropagation, based on this dual-component loss function, the joint update of parameters for both the server sub--network ($f_{\text{S}}$) and the edge device sub--network ($f_{\text{E}}$), along with the FC layer, is facilitated. 

Although out of the scope of this paper, the early-exit case requires additional decision-making if the IoT device makes local decision or transmits the intermediate representation to the server. Unlike making device-side decision, we propose to delegate this task to the server side. Namely, based on the estimated accuracy (prediction MSE), the server may decide to ask the device to forward its intermediate representations (in case of low MSE at the server), or to make the decision locally (if the server-side MSE is exceedingly high). This means that the edge devices receive periodic updates regarding their operation mode from the server side.  

%\subsubsection{Early--Exit with Flexible Loss Function}
%\label{early_weight}
The loss function, defined in Eq. (\ref{loss_early}), represents a simple sum of two distinct loss functions, each contributing equally to the training process and overall system performances. Alternatively, we may have prior knowledge about the system's operating conditions or specific requirements regarding desirable latency or communication overhead. In such cases, guiding the edge sub--network toward optimal decisions based on this prior knowledge would be beneficial. In our previous work \cite{ninkovic_2021, ninkovic_2023}, we considered more general case of \textit{compound loss function}, which exerts influence on the training process by assigning appropriate weights to different loss components. As a consequence, one may effectively balance performance among a diverse range of users. Inspired by this approach, a similar weighted sum generalization can be applied to Eq. (\ref{loss_early}) in order to favor certain decisions made by the server (early or final exit):
\begin{align}
\label{loss_early_weight}
    \mathcal{L}_{ee}=\lambda\mathcal{L}(\hat{y}, y)+(1-\lambda)\mathcal{L}(\hat{y}_{ee}, y),
\end{align}
where $\lambda$ denotes weight parameter and $\lambda\in[0,1]$. While a simple alteration of the loss function, the suggested approach introduces a 'knob' (parameter $\lambda$) that can be adjusted to shape the desired behavior of the system. Moreover, it can offer a flexible trade-off between early--exit—significant in critical applications where latency should be reduced and predictions should be obtained as soon as possible—and full system performance, where errors should be minimized.

\subsection{COMSPLIT with Heterogeneous IoT Devices}
\label{COMSPLIT_hetIoT}
IoT networks usually consist of devices boasting diverse computational capabilities (from simple sensors to cameras), enabling them to collect and analyze data originating from various source modalities (time series data, images, etc.). To achieve optimal data preprocessing, COMSPLIT introduces a novel pipeline, which employs different sub--networks on edge devices, each tailored to the complexity of the collected data and the computational capability of the device (Section \ref{hetIoT}).

Although traditional IoT networks consist of numerous devices, for simplicity, we will examine the scenario depicted in Fig. \ref{fig_hIoT}, where two edge devices communicate with a server, and $\mathcal{C}(f_{\text{E}_1})<\mathcal{C}(f_{\text{E}_2)}$. The principles discussed can easily be extended to accommodate larger IoT networks.  

In the proposed approach, each edge device has its local dataset $\mathcal{D}_i$, where $i\in\{1,2\}$, consisting of $(\boldsymbol{x}_i, y_i)$ pairs (Section \ref{hetIoT}). Throughout the offline split learning activation (forward) propagation phase (illustrated by the blue arrows in Fig. \ref{fig_hIoT}), each individual edge device generates its own intermediate representation using its own sub--network ($\boldsymbol{z}_i=f_{\text{E}_i}(\boldsymbol{x}_i), i\in\{1,2\}$). The intermediate representations are transmitted to the server side (using time--division scheduled access, Section \ref{hetIoT}), where they are concatenated to form $\boldsymbol{Z}=(\boldsymbol{z}_1, \boldsymbol{z}_2)$. After transmission through erasure channel, at the server side, the distorted $\mathcal{W}(\boldsymbol{Z})=\hat{\boldsymbol{Z}}=(\hat{\boldsymbol{z}}_1, \hat{\boldsymbol{z}}_2)$ is processed by the sub--network $f_{\text{S}}$, resulting in $(\hat{y}_1, \hat{y}_2) = f_{\text{S}}(\hat{\boldsymbol{Z}})=f_{\text{S}}(\hat{\boldsymbol{z}}_1, \hat{\boldsymbol{z}}_2)$. Based on the output predictions, the loss function (Eq. (\ref{loss_het})) is calculated on the server side, and the parameters of $(\{f_{\text{E}_i}\}_{i=1,2}, f_{\text{S}})$ are jointly optimized and updated through backpropagation (Fig. \ref{fig_hIoT}, red arrows). %Similar weighted approach, introduced in Section \ref{early_weight} can be applied here, but we must be fair and said that if IoT network comprises $C$ devices, then $C-1$ weight parameters inside loss function must be optimized. 

The training method described above enhances the server sub--network's resilience to fluctuating channel conditions in the online split inference phase,  simultaneously empowering edge devices to optimize intermediate representations. Additionally, it facilitates adaptive resource allocation within edge sub-networks, enabling devices to optimize overall system performance by adjusting the appropriate level of computational effort applied during preprocessing.

\section{Performance Evaluation}
In this section, for the above--described scenarios, we evaluate the system performance during the online inference phase and compare it with conventional split learning/inference scheme\footnotemark[3] \footnotetext[3]{Given the lack of direct references in the literature for SL approaches tailored to time series processing, our study uses a communication-agnostic vanilla SL scheme as a reference to underscore the benefits of COMSPLIT.}. We are particularly interested in the quality of inference when channel conditions deviate from the ones that were observed during the training phase. Two scenarios are elaborated in the following performance evaluation. The first and more general exploits time series data with the goal of exploring the applicability of COMSPLIT design and its potential deployment across generic IoT applications. The second and more specific targets evaluation of COMSPLIT design as part of a water quality monitoring system on the real-world data collected through practical scenario deployment. This analysis provides insights into how robustly the system performs under real--world conditions, specifically, its ability to maintain inference quality amidst changing channel conditions. 

\subsection{COMSPLIT Experimental Setup}
\subsubsection{\textit{Use Case 1: Generic Time-Series IoT System}}\label{scen1}
In the first use case, the extensive examination of COMSPLIT design behavior is conducted using a generic dataset that emulates an IoT system with time series data.  The findings from this broad analysis are then verified in a more specific IoT application. More precisely, COMSPLIT is evaluated on the \textit{Amazon Stock Data} dataset, which consists of 6516 rows. Each row represents data collected over one day and includes seven different pieces of information (stock prices): 'Date', 'Open', 'High', 'Low', 'Close', and 'Volume'. Throughout all conducted experiments, our objective is to forecast the 'Open' stock value for the current day by utilizing the 'Close' value from the preceding 30 days. While this dataset is not directly related to an IoT scenario, it encapsulates the behavior of time series data. The proposed approach can be easily extended to other time-series IoT scenarios, since the prediction pipeline remains the same, as will be shown in the Section \ref{scenWater}.

To prepare the time series data for training and testing, we define several preprocessing steps. Initially, we skip the first 30 rows. In other words, the 'Close' values from these initial rows serve as input for the first instance in our dataset. This approach ensures a consistent window of data for each prediction, utilizing the 'Close' values from the preceding 30 days as features. As a result, the dataset consists of 6486 instances, with 60\%\ (3892 instances) used for training, 10\%\ (648 instances) for validation, and 30\%\ (1946 instances) for testing. Each instance includes one label ('Open' value for the current day) and 30 features ('Close' values for the preceding 30 days). Moreover, as the utilized dataset collects information about Amazon stock prices for almost 18 years, there is significant variability within the data. To address this variability and ensure stable model training, we normalize the data to values between -1 and 1. 

\subsubsection{Use Case 2: Water Quality Monitoring IoT System}
\label{scenWater}
\begin{figure}
	\centering
	\includegraphics[width=0.85\linewidth]{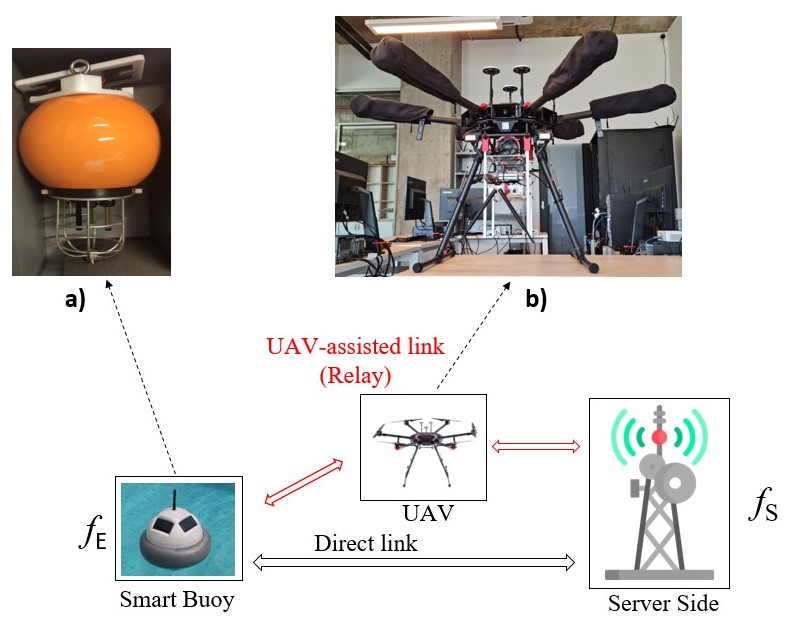}
  \vspace{-3mm}
	\caption[figure caption]{Use case 2: Water quality monitoring IoT system - Initial setup with a) smart buoy and b) UAV equipped with communication equipment\footnotemark[4]}
	\label{fig_dunav_scen}
\end{figure}

The second use case marks an important advancement in deploying the proposed COMSPLIT design principles in a real--world setting. The scenario evaluates COMSPLIT design in developing a SL-based solution for a water quality monitoring system. In terms of IoT system architecture, as presented in Fig. \ref{fig_dunav_scen}, a smart buoy, which integrates all sensors and communication equipment, communicates with a conventional stationary server. Moreover, as part of our ongoing, we explore the potential of communication with an unmanned aerial vehicle (UAV), which could serve both as a server and as a UAV--assisted (relay) link \cite{sabze_2022} (Fig. \ref{fig_dunav_scen}, red arrows). We note that this system, as illustrated in Fig. \ref{fig_dunav_scen}a (smart buoy) and \ref{fig_dunav_scen}b (UAV equipped with communication equipment), is integrated into UAV--assisted relaying platform \cite{ninkovic_2024}.

The goal of this case is to investigate how the proposed design can handle real-world data collected from sensors deployed in operational systems. These sensors can exhibit imperfections such as calibration errors, drift, and environmental noise, which can affect the data and assist in evaluating the robustness of the COMSPLIT design. The dataset used in this case is particularly well-suited for the water quality monitoring problem of interest, focusing on the Danube river near the city of Novi Sad (Serbia). It contains 3,264 data points, with 70\%\ allocated for training and 30\%\ for testing. Each data point represents a daily measurement of 
%published by the Republic Hydrometeorological Service of Serbia, spanning from November 2013 to October 2022
critical water quality parameters: temperature, pH value, electrical conductivity, dissolved oxygen, oxygen saturation, ammonium, and nitrite.

In this setup, dissolved oxygen is predicted by using  the last 30 measurements (covering the previous 30 days). After preprocessing and converting the data to a time series format, each instance in the dataset consists of 30 features (30 previous dissolved oxygen measurements) and one label (representing the dissolved oxygen level for the current day). Similar to the previous case, the data is normalized to a range of -1 to 1 to handle the varying scales of the parameters. Although our setup and equipment allow us to track all eight parameters for 30 days, resource-constrained IoT devices typically face limitations in data storage. Therefore, our goal was to design the experiment to correspond to real-world practical constraints. While multivariate distributed learning is beyond the scope of this work, it will be addressed in future research. 

\footnotetext[4]{UAV--assisted link (realy) is beyond scope of this paper, it will be included in future work.}

\subsection{COMSPLIT Training Procedure}
\label{training_COMSPLIT}

Although the training procedure is tailored to each scenario (described in Section \ref{model}), there are several commonalities. Firstly, across all experiments, we optimize the appropriate loss function using stochastic gradient descent with the Adam optimizer \cite{adam} on a batch--by--batch basis. The learning rate is set to $\alpha = 0.001$, $\beta_1 = 0.9$, and $\beta_2 = 0.999$, while the batch size is set to 64. Following a similar approach to \cite{OShea_2017}, we keep the channel conditions constant during training (i.e., erasure probability for the erasure channel or noise variances for the AWGN channel), which can be considered as a neural network hyperparameter. As a result, the robustness of COMSPLIT is being thoroughly tested, pushing its limits by evaluating this approach in channel conditions that may significantly differ from those encountered during training. The analysis in Section \ref{results} will examine how the introduction of various channel conditions during the training process affects the overall system performances. The training procedure and hyperparameters remain consistent across all experiments related to the use cases defined in the previous subsection. The notation and training/testing parameter descriptions for: a) Erasure Channel, b) AWGN Channel, c) Early-Exit, and d) Heterogeneous Devices scenario, are given in Table \ref{table_training}.

\begin{table}[tbhp]
\caption{red}{
Notation \&\ Training/Testing Parameter Descriptions for Scenarios: a) Erasure Channel, b) AWGN Channel, c) Early-Exit, and d) Heterogeneous Devices}
\begin{center}
\begin{tabular}{|c|c|c|c|}
\hline
Scenario(s)&Parameter Notation&Parameter Description\\%&\textbf{Parameter Value}\\
\hline
A), B), C), D) &$N$& \makecell{Raw data $\boldsymbol{x}$ input size \\ (D) - Per device)}\\% &30\\ 
\hline
A), B), C), D) &$M$&\makecell{ $\boldsymbol{z}$ length \\ (D) - Per device)} \\%&\makecell{4 - Dotted red curve (Fig. \ref{Fig_comp})\\
%7 - Dashed red curve (Fig. \ref{Fig_comp}) \\ 
%10 - All solid curves \\ (Figs. \ref{Fig_erasure}, \ref{Fig_comp}, \ref{Fig_awgn}, \ref{Fig_awgn_1}, \ref{Fig_hIoT}, \ref{Fig_erasure_case2}, \ref{Fig_awgn_case2}, \ref{Fig_hIoT_2});\\Dashed curves (Figs. \ref{Fig_hIoT} and \ref{Fig_hIoT_2})} \\
\hline
A), B), C), D)&$H$&\makecell{LSTM hidden layer size\\ (D) - Per device)}\\%&\makecell{H=M:\\4 - Dotted red curve (Fig. \ref{Fig_comp})\\
%7 - Dashed red curve (Fig. \ref{Fig_comp}) \\ 
%10 - All solid curves (Figs. \ref{Fig_erasure}, \ref{Fig_comp}, \ref{Fig_awgn}, \ref{Fig_awgn_1}, \ref{Fig_hIoT}, \ref{Fig_erasure_case2}, \ref{Fig_awgn_case2}, \ref{Fig_hIoT_2});\\Dashed curves (Figs. \ref{Fig_hIoT} and \ref{Fig_hIoT_2})}\\
\hline
A), D)&$p_{tr}$& Training erasure probability \\%&\makecell{0 - Blue solid curve (Figs. \ref{Fig_erasure} and \ref{Fig_erasure_case2})\\
%0.1 - Red solid curve (Figs. \ref{Fig_erasure}, \ref{Fig_hIoT}, \ref{Fig_erasure_case2}, and \ref{Fig_hIoT_2});\\Red dashed curve (Figs. \ref{Fig_hIoT} and \ref{Fig_hIoT_2})\\
%All experiments presented in Fig. \ref{Fig_comp} \\
%0.2 - Green solid curve (Figs. \ref{Fig_erasure} and \ref{Fig_erasure_case2}) \\
%0.3 - Purple solid curve (Figs. \ref{Fig_erasure} and \ref{Fig_erasure_case2}) \\
%0.4 - Brown solid curve (Figs. \ref{Fig_erasure} and \ref{Fig_erasure_case2}) \\
%0.5 - Black solid curve (Figs. \ref{Fig_erasure}, \ref{Fig_hIoT}, \ref{Fig_erasure_case2}, and \ref{Fig_hIoT_2});\\ Black dashed curve (Figs. \ref{Fig_hIoT} and \ref{Fig_hIoT_2})} \\
\hline
A), D)&$p$&Testing erasure probability\\%$\{0, 0.1, \dots, 0.9\}\\%$ (Figs. \ref{Fig_erasure}, \ref{Fig_comp}, \ref{Fig_hIoT}, \ref{Fig_erasure_case2}, and \ref{Fig_hIoT_2})\\
\hline
A)& $CR$& Compression rate\\%& \makecell{$CR=N/M$= \\7.5 - Red dotted curve (Fig. \ref{Fig_comp})\\ 4.285 - Red dashed curve (Fig. \ref{Fig_comp})\\3 - Red solid curve (Fig. \ref{Fig_comp})}\\
\hline
B)& \makecell{ $SNR_1$ \\ $SNR_2=$\\$SNR_1-5$ dB}&\makecell{Training/Testing $SNR$: \\
$SNR_1$ - Non deep \\fade symbols\\
$SNR_2$ - Deep fade\\ symbols}\\%& \makecell{No channel - Blue solid curve (Figs. \ref{Fig_awgn} and \ref{Fig_awgn_case2})\\
%$SNR_1=-5$ dB - Red solid curve (Figs. \ref{Fig_awgn} and \ref{Fig_awgn_case2})\\
%$SNR_1=0$ dB - Green solid curve (Figs. \ref{Fig_awgn} and \ref{Fig_awgn_case2})\\
%$SNR_1=5$ dB - Purple solid curve (Figs. \ref{Fig_awgn} and \ref{Fig_awgn_case2})\\
%$SNR_1=10$ dB - Black solid curve (Figs. \ref{Fig_awgn} and \ref{Fig_awgn_case2})}\\
\hline
B)&\makecell{$M_1$ \\ $M_2=M-M_1$}& \makecell{$M_1$ - No. of \\non deep fade symbols \\ $M_2$ - No. of \\deep fade symbols }\\%& \makecell{$M_1=5$ - All solid curves (Figs. \ref{Fig_awgn} and \ref{Fig_awgn_case2});\\ Green solid curve (Fig. \ref{Fig_awgn_1})\\ $M_1=3$ - Black solid curve (Fig. \ref{Fig_awgn_1}) \\ $M_1=7$ - Red solid curve (Fig. \ref{Fig_awgn_1})}\\
\hline
C)&$\lambda$& \makecell{Weight parameter in early--exit\\ loss function (Eq. \ref{loss_early_weight})}\\%& 0.5 (Figs. \ref{Fig_erasure}, \ref{Fig_awgn}, \ref{Fig_erasure_case2}, \ref{Fig_awgn_case2})\\
\hline
D)&\makecell{$MSE_1$\\$MSE_2$}&\makecell{$MSE_1$ - MSE for device 1 \\(sub--network $f_{\text{E}_1}$ in Fig. \ref{fig_hIoT})\\$MSE_2$ - MSE for device 2 \\(sub--network $f_{\text{E}_2}$ in Fig. \ref{fig_hIoT})}\\%& \makecell{$MSE_1$ - Blue/Red/Black solid curve \\(Figs. \ref{Fig_hIoT} and \ref{Fig_hIoT_2}) \\$MSE_2$ - Blue/Red/Black dashed curve \\(Figs. \ref{Fig_hIoT} and \ref{Fig_hIoT_2})}\\
\hline
\end{tabular}
\label{table_training}
\end{center}
\end{table}

In the first three scenarios (Fig. \ref{fig_scen}), the overall system consists of an edge device and a server and the same neural network architecture is used. More precisely, we employ an LSTM neural network comprising 3 LSTM layers and one FC layer. The division point lies after the initial layer, meaning that the edge device handles one LSTM layer, while the remaining ones are executed on the server side. Each LSTM layer maintains a consistent hidden layer size $H$, ensuring that the output of each layer contains the same number of features. Consequently, the output of the first LSTM layer is an intermediate representation $\boldsymbol{z}$ (i.e., $H=M$) (which can be understood as a compressed representation of the raw data, Section \ref{SL}). The size of this representation, which defines the compression rate, significantly impacts both the training process and the overall system performance, which is analyzed in Section \ref{results}. The FC layer, whose output represents the estimated value, is a conventional linear layer consisting of $M$ neurons. For the early-exit strategy, an additional FC layer with $M$ neurons is attached to the LSTM layer at the edge device, enabling local predictions.

In the heterogeneous IoT scenario, consisting of two edge devices with different computational capabilities and a server (Fig. \ref{fig_hIoT}), the training procedure is slightly adjusted. Namely, the initial batch size is divided into two equal parts, each processed by a different edge device, and instances within each batch are shuffled during the training process. The sub--network configuration differs between edge devices and the server: edge devices incorporate one and two LSTM layers, respectively, while the server's sub-network maintains a consistent structure across both edge devices, comprising 3 LSTM layers and one FC layer (Fig. \ref{fig_hIoT}). The rest of the procedure aligns with that described for the initial three scenarios, with a consistent hidden size ($H=M$) across all LSTM layers and an identical number of neurons in the FC layer.   

\subsection{Numerical Results - Use Case 1 (Generic Time-Series IoT System)}
\label{results}
The goal of this work is to devise a design of split neural network architecture which will remain robust to varying channel conditions during the (online) inference phase. More precisely, after providing comprehensive analysis of the COMSPLIT performance, we suggest a suitable strategy for the communication--aware neural network design for generic IoT systems that utilize time series data.  

\subsubsection{Erasure Channel}
\label{erasure_perfm}

\begin{figure}[t]
	\begin{tikzpicture}[spy using outlines=
{rectangle, red, magnification=3.2, line width=2, connect spies}]
  	\begin{semilogyaxis}[width=1\columnwidth, height=7.5cm, 
	legend style={at={(0.59,0.98)}, anchor= north,font=\scriptsize, legend style={nodes={scale=0.9, transform shape}}},
   	legend cell align={left},
	legend columns=3,   	 
   	x tick label style={/pgf/number format/.cd,
   	set thousands separator={},fixed},
   	y tick label style={/pgf/number format/.cd,fixed, precision=2, /tikz/.cd},
   	xlabel={$p$},
   	ylabel={MSE},
   	label style={font=\footnotesize},
   	grid=major,   	
   	xmin = 0, xmax = 0.9,
   	ymin=0.0007, ymax=1,
   	line width=0.8pt,
   	tick label style={font=\footnotesize},]
   	\addplot[blue, mark=square] 
   	table [x={x}, y={y}] {./Figs./dropout/ir_10_0};
   	\addlegendentry{No channel}
   	\addplot[red, mark=x, mark options={solid}] 
   	table [x={x}, y={y}] {./Figs./dropout/ir_10_01};
   	\addlegendentry{$p_{tr}=0.1$}
   	
   	\addplot[green, mark=triangle, mark options={solid}] 
   	table [x={x}, y={y}] {./Figs./dropout/ir_10_02};
   	\addlegendentry{$p_{tr}=0.2$}
   	
   	\addplot[purple, mark=o, mark options={solid}] 
   	table [x={x}, y={y}] {./Figs./dropout/ir_10_03};
   	\addlegendentry{$p_{tr}=0.3$}
   	\addplot[brown, mark=star, mark options={solid}] 
   	table [x={x}, y={y}] {./Figs./dropout/ir_10_04}; 	\addlegendentry{$p_{tr}=0.4$}
   	\addplot[black, mark=diamond, mark options={solid}] 
   	table [x={x}, y={y}] {./Figs./dropout/ir_10_05};
        \addlegendentry{$p_{tr}=0.5$}
        \addplot[line width=0.6mm, dashed, mark=none, red, samples=2]{0.00194};
  \addlegendentry{Early--exit}
\coordinate (spypoint) at (axis cs:0.,0.001);
\coordinate (spyviewer) at (axis cs:0.08, 0.09);
\spy[width=1cm,height=2.7cm] on (spypoint) in node [fill=white] at (spyviewer);
 	\end{semilogyaxis}
	\end{tikzpicture}
	\vspace*{-5mm}
	\caption{Erasure channel: MSE versus symbol erasure probability ($p$) performances for various $p_{tr}$ values introduced during training and early--exit strategy - Use Case 1.}
	\label{Fig_erasure}
\end{figure}
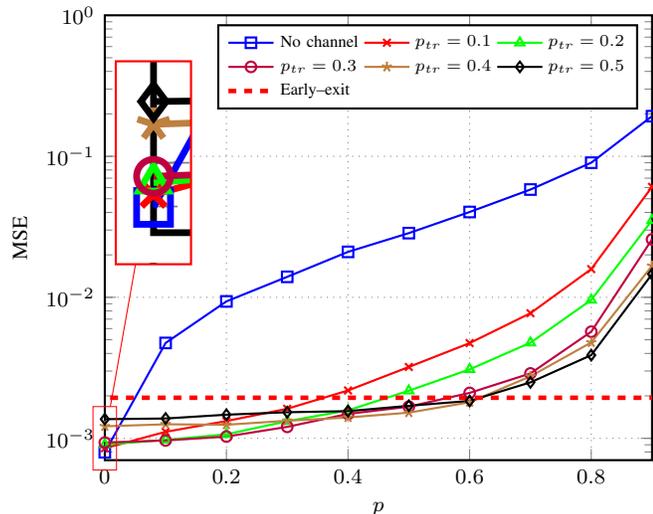

In Fig. \ref{Fig_erasure}, we assess the performance of the proposed COMSPLIT approach in terms of prediction mean--squared error (MSE) and compare it with a conventional split learning/inference system trained without channel impairments between the edge device and server (ideal channel) when $H=M=10$. Throughout the training phase, we uphold fixed channel conditions with a symbol erasure probability denoted as $p_{tr}$, while testing is done across a range of symbol erasure probabilities denoted as $p$ (Table \ref{table_training}). 
The comparison reveals a stark performance contrast: while the conventional split learning/inference approach experiences notable degradation as soon as the channel begins to affect the intermediate representation (illustrated by the blue curve in Fig. \ref{Fig_erasure}), the proposed approach exhibits graceful MSE degradation demonstrating high resilience to diverse channel conditions encountered during testing.
Of particular significance is the impact of varying channel conditions introduced during the training phase. Specifically, under good channel conditions ($p=0$), optimal performance favors the conventional approach and approaches with $p_{tr}=0.1-0.2$, whereas the poorest results are manifested for $p_{tr}=0.5$ (as delineated by the red magnification in Fig. \ref{Fig_erasure}). Conversely, in scenarios where the channel exerts a substantial influence ($p=0.7-0.9$), superior performance aligns with $p_{tr}=0.5$, starkly contrasting the significant performance deterioration observed in the conventional approach.

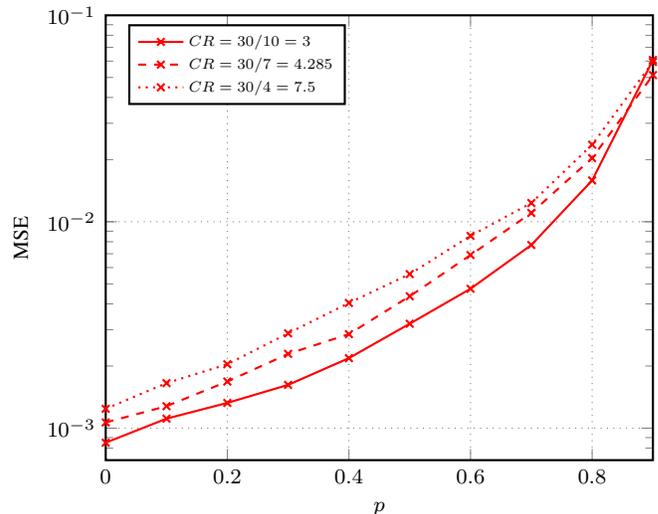
\begin{figure}[t]
	\begin{tikzpicture}[spy using outlines=
{rectangle, magnification=4, connect spies}]
  	\begin{semilogyaxis}[width=1\columnwidth, height=7.5cm, 
	legend style={at={(0.24
 ,0.98)}, anchor= north,font=\scriptsize, legend style={nodes={scale=0.8, transform shape}}},
   	legend cell align={left},
	legend columns=1,   	 
   	x tick label style={/pgf/number format/.cd,
   	set thousands separator={},fixed},
   	y tick label style={/pgf/number format/.cd,fixed, precision=2, /tikz/.cd},
   	xlabel={$p$},
   	ylabel={MSE},
   	label style={font=\footnotesize},
   	grid=major,   	
   	xmin = 0, xmax = 0.9,
   	ymin=0.0007, ymax=0.1,
   	line width=0.8pt,
   	tick label style={font=\footnotesize},]
   	\addplot[red, mark=x] 
   	table [x={x}, y={y}] {./Figs./dropout/ir_10_01};
   	\addlegendentry{$CR=30/10=3$}
   	\addplot[dashed, red, mark=x, mark options={solid}] 
   	table [x={x}, y={y}] {./Figs./dropout/ir_7_01};
   	\addlegendentry{$CR=30/7=4.285$}
   	
   	\addplot[dotted, red, mark=x, mark options={solid}] 
   	table [x={x}, y={y}] {./Figs./dropout/ir_4_01};
   	\addlegendentry{$CR=30/4=7.5$}
 	\end{semilogyaxis}
	\end{tikzpicture}
	\vspace*{-5mm}
	\caption{MSE versus erasure probability $p$ for different compression rates obtained at $p_{tr}=0.1$ - Use Case 1.}
	\label{Fig_comp}
\end{figure}

The compression rate ($CR=N/M=30/M$) plays a significant role in determining the overall system performance, impacting not only MSE but also various other performance aspects (such as the communication overhead). In Fig. \ref{Fig_comp}, we present a comparison of prediction MSE achieved with the COMSPLIT approach trained at $p_{tr}=0.1$ and tested across the range of values $p$, considering three different compression rates, when $H=M=\{4, 7, 10\}$. As expected, extending the representation length results in enhanced MSE performance.  Importantly, even with a high compression rate ($H=M=4$), we observe decent performance levels. Furthermore, increased compression rates not only reduce the communication overhead but introduce an additional adaptation parameter to the proposed approach. However, detailed analysis of compression rate influence on system design exceeds the scope of this work. 

\subsubsection{AWGN Channel}
\label{awgn_perf}
\begin{figure}[t]
	\begin{tikzpicture}[spy using outlines=
{rectangle, magnification=4, connect spies}]
  	\begin{semilogyaxis}[width=1\columnwidth, height=7.5cm, 
	legend style={at={(0.553,0.98)}, anchor= north,font=\scriptsize, legend style={nodes={scale=0.8, transform shape}}},
   	legend cell align={left},
	legend columns=2,   	 
   	x tick label style={/pgf/number format/.cd,
   	set thousands separator={},fixed},
   	y tick label style={/pgf/number format/.cd,fixed, precision=2, /tikz/.cd},
   	xlabel={$SNR_1$ [dB]},
   	ylabel={MSE},
   	label style={font=\footnotesize},
   	grid=major,   	
   	xmin = -10, xmax = 10,
   	ymin=0.0005, ymax=1,
   	line width=0.8pt,
   	tick label style={font=\footnotesize},]
   	\addplot[blue, mark=square] 
   	table [x={x}, y={y}] {./Figs./awgn/7_3_no_channel};
   	\addlegendentry{Training-No channel}
   	\addplot[red, mark=x, mark options={solid}] 
   	table [x={x}, y={y}] {./Figs./awgn/5_5_channel_m5_m10};
   	\addlegendentry{Training - $SNR_1=-5$ dB}
   	
   	\addplot[green, mark=triangle, mark options={solid}] 
   	table [x={x}, y={y}] {./Figs./awgn/5_5_channel_0_m5};
   	\addlegendentry{Training - $SNR_1=0$ dB}
   	
   	\addplot[purple, mark=o, mark options={solid}] 
   	table [x={x}, y={y}] {./Figs./awgn/5_5_channel_5_0};
   	\addlegendentry{Training - $SNR_1=5$ dB}
   	\addplot[black, mark=star, mark options={solid}] 
   	table [x={x}, y={y}] {./Figs./awgn/5_5_channel_10_5}; 	\addlegendentry{Training - $SNR_1=10$ dB}
     \addplot[dashed, red, mark=none, line width=0.6mm] 
   	table [x={x}, y={y}] {./Figs./awgn/early}; 	\addlegendentry{Early--exit}
  \addlegendentry{Early--Exit}
   	
 	\end{semilogyaxis}
	\end{tikzpicture}
	\vspace*{-8mm}
	\caption{AWGN channel: MSE versus $SNR$ performances for various $SNR_1$ and $SNR_2$ values introduced during training and early--exit strategy - Use Case 1.}
	\label{Fig_awgn}
\end{figure}
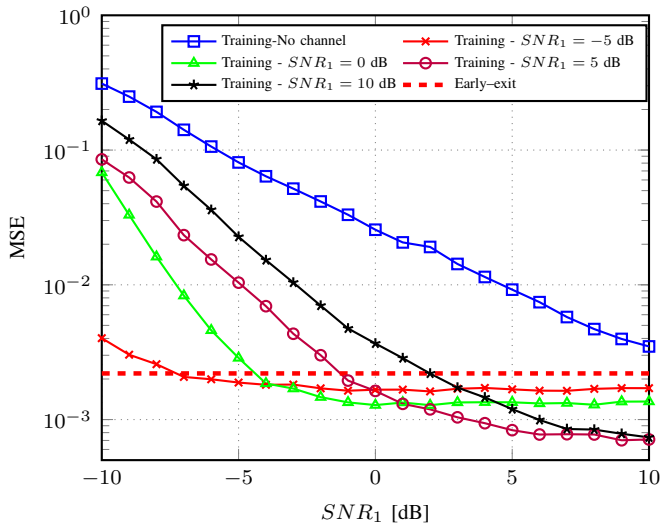

We examine the AWGN channel model (Section \ref{AWGN}) where the number of symbols experiencing different noise variances is equal, i.e., $M_1=M_2=5$ ($H=M=10$). To analyze how prediction MSE performance depends on the AWGN channel configuration presented during the training phase, we define four different sub--scenarios where $SNR_1 \in \{-5, 0, 5, 10\}$ dB, and $SNR_2$ is set to $SNR_1-5$ dB. In Fig. \ref{Fig_awgn}, prediction MSE versus $SNR_1$ is compared for the 4 sub--scenarios, including the case of conventional split learning/inference approach (trained on the ideal channel). It is noteworthy that the conventional approach exhibits significant (order of magnitude) performance degradation, even at high $SNR_1$ values, while COMSPLIT demonstrates graceful improvement as channel conditions improve. 

 We are particularly interested in analyzing the impact of varying SNR values introduced during training on the overall system performance. If a neural network model is trained on extremely poor channel conditions ($SNR_1=-5$ dB, $SNR_2=-10$ dB, depicted by the red solid curve in Fig. \ref{Fig_awgn}), performance optimization occurs for $SNR_1$ values between $-10$ dB and $-5$ dB, with no improvement observed for better channel conditions. On the other hand, a model trained with $SNR_1=5$ dB and $SNR_2=0$ dB (depicted by the purple curve in Fig. \ref{Fig_awgn}) exhibits notable enhancement in performance as $SNR$ values improve. This finding suggests a possibility of pre-training several neural network models for different SNR ranges during the offline training phase. Subsequently, the server can dynamically select the most suitable model based on the estimated channel conditions. Finally, comparing Figs. \ref{Fig_erasure} and \ref{Fig_awgn}, it is interesting to note that the system performance for AWGN trained and tested at high SNR value ($SNR_1=10$ dB), for which prediction MSE equals $7\times10^{-4}$, slightly exceeds the system performance of noiseless channel ($p=0$), where MSE equals $8\times10^{-4}$. This improvement can be attributed to the regularization introduced by the noise layer (i.e., low-noise AWGN channel) during the training process.

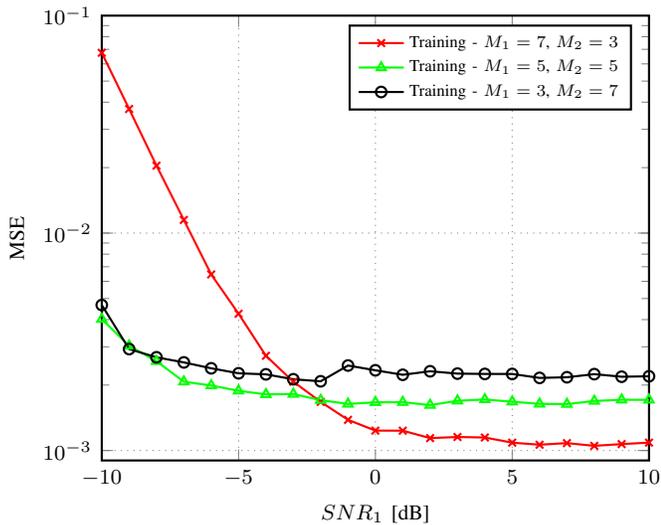
\begin{figure}[t]
	\begin{tikzpicture}[spy using outlines=
{rectangle, magnification=4, connect spies}]
  	\begin{semilogyaxis}[width=1\columnwidth, height=7.5cm, 
	legend style={at={(0.71,0.98)}, anchor= north,font=\scriptsize, legend style={nodes={scale=0.9, transform shape}}},
   	legend cell align={left},
	legend columns=1,   	 
   	x tick label style={/pgf/number format/.cd,
   	set thousands separator={},fixed},
   	y tick label style={/pgf/number format/.cd,fixed, precision=2, /tikz/.cd},
   	xlabel={$SNR_1$ [dB]},
   	ylabel={MSE},
   	label style={font=\footnotesize},
   	grid=major,   	
   	xmin = -10, xmax = 10,
   	ymin=0.0009, ymax=0.1,
   	line width=0.8pt,
   	tick label style={font=\footnotesize},]
   	%\addplot[blue, mark=square] 
   	%table [x={x}, y={y}] {./Figs./awgn/7_3_no_channel};
   	%\addlegendentry{Training-No channel}
   	\addplot[red, mark=x, mark options={solid}] 
   	table [x={x}, y={y}] {./Figs./awgn/7_3_channel_0_m5};
   	\addlegendentry{Training  - $M_1=7$, $M_2=3$ }
   	
   	\addplot[green, mark=triangle, mark options={solid}] 
   	table [x={x}, y={y}] {./Figs./awgn/5_5_channel_m5_m10};
   	\addlegendentry{Training - $M_1=5$, $M_2=5$}
   	
   	\addplot[black, mark=o, mark options={solid}] 
   	table [x={x}, y={y}] {./Figs./awgn/3_7_channel_m5_m10};
   	\addlegendentry{Training - $M_1=3$, $M_2=7$}
   	
 	\end{semilogyaxis}
	\end{tikzpicture}
	\vspace*{-8mm}
	\caption{AWGN channel: MSE versus $SNR$ performances for various $M_1$ and $M_2$ values trained at $SNR_1=-5$ dB - Use Case 1.}
	\label{Fig_awgn_1}
\end{figure}

Recalling Section \ref{AWGN}, the number and positions of symbols affected by higher noise variance (deep fades) is randomly selected. In Fig. \ref{Fig_awgn_1}, we present prediction MSE performance versus $SNR_1$ for $M_1\in \{7, 5, 3\}$ and for training $SNR_1=-5$ dB. We observe that for $M_1=7$, the model enhances its performance as $SNR_1$ increases. Conversely, for the remaining two values of $M_1$, achieving improvements with increased $SNR_1$ proves more challenging due to adverse effect of $M_2$ symbols. Here, the model optimizes its performance based on the trained $SNR_1$ value, without improvements as the channel conditions improve.

\subsubsection{Early--Exit Strategy}
\label{early_perf}
In Figs. \ref{Fig_erasure} and \ref{Fig_awgn}, alongside the results obtained with the erasure and AWGN channels, respectively, we present the outcomes of the early-exit strategy (horizontal dashed red line) for $\lambda=0.5$ (Eq. \ref{loss_early_weight}). Note that the early-exit performance is achieved without transmitting intermediate representations $\boldsymbol{z}$ to the server, as the decision is made locally at the edge. Consequently, this performance remains unaffected by channel conditions and is represented by a single value (i.e., horizontal line). The results indicate that this strategy holds significant potential for scenarios of poor channel conditions between edge devices and servers. In such situations, when symbol erasure probability $p>0.4$ (Fig. \ref{Fig_erasure}) or $SNR_1<0$ db (Fig. \ref{Fig_awgn}), the preferable choice is to make local predictions, avoiding transmission of the intermediate representation. 

\begin{table}[tbhp]
\caption{ Earl--exit versus full performances regarding loss function parameter $\lambda$.}
\begin{center}
\begin{tabular}{|c|c|c|}
\hline
$\lambda$&Early--exit MSE&Full system MSE ($p_{tr}=0.1$) \\
\hline 
0.1&0.00102&0.00093\\
\hline
0.5&0.00194&0.00085\\
\hline
0.9&0.004&0.00052\\
\hline
\end{tabular}
\label{table_early}
\end{center}
\end{table}

As noted in \cite{yankowski_2023}, this approach significantly reduces the latency, and could be further enhanced by adjusting the parameter $\lambda$ in the weighted loss function (Eq. \ref{loss_early_weight}). Specifically, a closer examination of Table \ref{table_early} provides insights into the influence of $\lambda$ on system performance. It reveals that when the loss function favors the early-exit strategy, the MSE obtained with local decisions is comparable to the ones at the output of the server (trained at $p_{tr}=0.1$). Thus by simply tuning of a single parameter $\lambda$, desirable performance levels can be achieved. This enables the server to dynamically select suitable strategy and pre--trained model based on the current channel conditions.

\subsubsection{Heterogeneous IoT Devices} 
\label{het_IoT_perf}
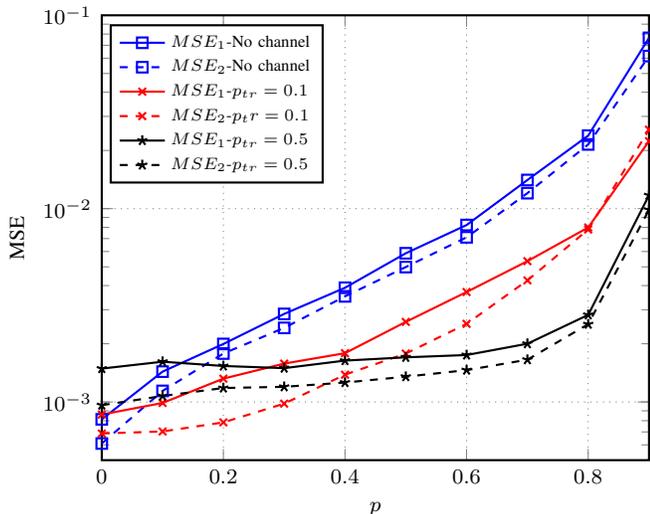
\begin{figure}[t]
	\begin{tikzpicture}[spy using outlines=
{rectangle, magnification=4, connect spies}]
  	\begin{semilogyaxis}[width=1\columnwidth, height=7.5cm, 
	legend style={at={(0.21,0.98)}, anchor= north,font=\scriptsize, legend style={nodes={scale=0.9, transform shape}}},
   	legend cell align={left},
	legend columns=1,   	 
   	x tick label style={/pgf/number format/.cd,
   	set thousands separator={},fixed},
   	y tick label style={/pgf/number format/.cd,fixed, precision=2, /tikz/.cd},
   	xlabel={$p$},
   	ylabel={MSE},
   	label style={font=\footnotesize},
   	grid=major,   	
   	xmin = 0, xmax = 0.9,
   	ymin=0.0005, ymax=0.1,
   	line width=0.8pt,
   	tick label style={font=\footnotesize},]
   	\addplot[blue, mark=square] 
   	table [x={x}, y={y}] {./Figs./hetIoT/mse_1_0};
   	\addlegendentry{$MSE_1$-No channel}
        \addplot[dashed, blue, mark=square, mark options=solid] 
   	table [x={x}, y={y}] {./Figs./hetIoT/mse_2_0};
   	\addlegendentry{$MSE_2$-No channel}

        \addplot[red, mark=x] 
   	table [x={x}, y={y}] {./Figs./hetIoT/mse_1_01};
   	\addlegendentry{$MSE_1$-$p_{tr}=0.1$}
        \addplot[dashed, red, mark=x, mark options=solid] 
   	table [x={x}, y={y}] {./Figs./hetIoT/mse_2_01};
   	\addlegendentry{$MSE_2$-$p_tr=0.1$}

        %\addplot[purple, mark=o] 
   	%table [x={x}, y={y}] {./Figs./hetIoT/mse_1_03};
   	%\addlegendentry{$MSE_1$-$p_{tr}=0.3$}
        %\addplot[dashed, purple, mark=o, mark options=solid] 
   	%table [x={x}, y={y}] {./Figs./hetIoT/mse_2_03};
   	%\addlegendentry{$MSE_2$-$p_{tr}=0.3$}

        \addplot[black, mark=star] 
   	table [x={x}, y={y}] {./Figs./hetIoT/mse_1_05};
   	\addlegendentry{$MSE_1$-$p_{tr}=0.5$}
        \addplot[dashed, black, mark=star, mark options=solid] 
   	table [x={x}, y={y}] {./Figs./hetIoT/mse_2_05};
   	\addlegendentry{$MSE_2$-$p_{tr}=0.5$}
    
 	\end{semilogyaxis}
	\end{tikzpicture}
	\vspace*{-5mm}
	\caption{Heterogeneous IoT Devices: $MSE$ versus symbol erasure probability $p$ performances for 2 devices ($\mathcal{C}(f_{\text{E}_1})<\mathcal{C}(f_{\text{E}_2})$) under different $p_{tr}$ - Use Case 1.}
	\label{Fig_hIoT}
\end{figure}

In this scenario, we examine the behavior of two devices that simultaneously communicate with a server over an erasure channel. Both devices produce individual intermediate representations of length $H=M=10$, independently experiencing the same channel conditions (as described in Section \ref{COMSPLIT_hetIoT}). In Fig. \ref{Fig_hIoT}, we present the prediction MSE when a dropout layer is incorporated into the training process ($MSE_1$ is connected with the edge sub--network $f_{\text{E}_1}$, and $MSE_2$ with sub--network $f_{\text{E}_2}$, see Fig. \ref{fig_hIoT}). We observe similar behavior as described in Section \ref{erasure_perfm}, where $p_{tr}$ has significant influence on the system performance. MSE values achieved with the COMSPLIT approach significantly surpass those of conventional split learning/inference algorithms for $p>0.1$. In all three cases depicted at Fig. \ref{Fig_hIoT}, the more capable edge device that handles two LSTM layers exhibits lower MSE values (dashed lines in Fig. \ref{Fig_hIoT}). With more powerful inference engine deployed at the edge device, the device is capable of providing accurate inference on more complex data (e.g., multi--modal data \cite{vep_2018}). We note that, as the central point of the system, the server could determine the optimal workload for each individual device based on desired performance metrics, channel conditions and the computational capabilities of these edge devices.

\subsection{Numerical Results - Use Case 2 (Water Quality Monitoring IoT System)}
\label{results_dunav}
Building on the findings from the generic IoT use case 1, we focus on a more specific use case 2 involving a water quality monitoring IoT system. The primary goal is to deploy a system that remains robust to channel perturbations while effectively handling real--world data and sensor impairments. This deployment leverages the proposed communication--aware neural network design, ensuring the system's reliability and accuracy in a practical setting.

\subsubsection{Erasure Channel}
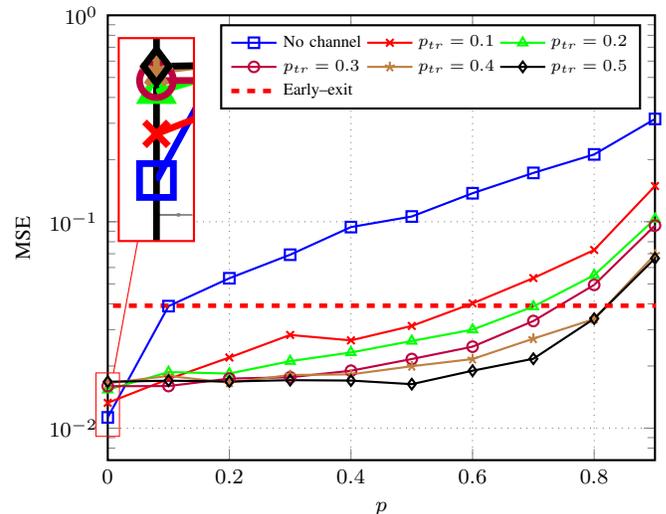
\begin{figure}[t]
	\begin{tikzpicture}[spy using outlines=
{rectangle, red, magnification=3.2, line width=2, connect spies}]
  	\begin{semilogyaxis}[width=1\columnwidth, height=7.5cm, 
	legend style={at={(0.59,0.98)}, anchor= north,font=\scriptsize, legend style={nodes={scale=0.9, transform shape}}},
   	legend cell align={left},
	legend columns=3,   	 
   	x tick label style={/pgf/number format/.cd,
   	set thousands separator={},fixed},
   	y tick label style={/pgf/number format/.cd,fixed, precision=2, /tikz/.cd},
   	xlabel={$p$},
   	ylabel={MSE},
   	label style={font=\footnotesize},
   	grid=major,   	
   	xmin = 0, xmax = 0.9,
   	ymin=0.007, ymax=1,
   	line width=0.8pt,
   	tick label style={font=\footnotesize},]
   	\addplot[blue, mark=square] 
   	table [x={x}, y={y}] {./Figs./Dunav/dropout/ir_10_0.txt};
   	\addlegendentry{No channel}
   	\addplot[red, mark=x, mark options={solid}] 
   	table [x={x}, y={y}] {./Figs./Dunav/dropout/ir_10_01.txt};
   	\addlegendentry{$p_{tr}=0.1$}
   	
   	\addplot[green, mark=triangle, mark options={solid}] 
   	table [x={x}, y={y}] {./Figs./Dunav/dropout/ir_10_02.txt};
   	\addlegendentry{$p_{tr}=0.2$}
   	
   	\addplot[purple, mark=o, mark options={solid}] 
   	table [x={x}, y={y}] {./Figs./Dunav/dropout/ir_10_03.txt};
   	\addlegendentry{$p_{tr}=0.3$}
   	\addplot[brown, mark=star, mark options={solid}] 
   	table [x={x}, y={y}] {./Figs./Dunav/dropout/ir_10_04.txt}; 	\addlegendentry{$p_{tr}=0.4$}
   	\addplot[black, mark=diamond, mark options={solid}] 
   	table [x={x}, y={y}] {./Figs./Dunav/dropout/ir_10_05.txt};
        \addlegendentry{$p_{tr}=0.5$}
        \addplot[line width=0.6mm, dashed, mark=none, red, samples=2]{0.0392};
  \addlegendentry{Early--exit}
\coordinate (spypoint) at (axis cs:0.,0.013);
\coordinate (spyviewer) at (axis cs:0.08, 0.25);
\spy[width=1cm,height=2.7cm] on (spypoint) in node [fill=white] at (spyviewer);
 	\end{semilogyaxis}
	\end{tikzpicture}
	\vspace*{-5mm}
	\caption{Erasure channel: MSE versus symbol erasure probability ($p$) performances for various $p_{tr}$ values introduced during training and early--exit strategy - Use Case 2.}
	\label{Fig_erasure_case2}
\end{figure}

Fig. \ref{Fig_erasure_case2} illustrates the prediction MSE performance of the IoT water monitoring system, designed and trained offline based on COMSPLIT principles.  The figure compares the performance of the online inference phase fed with real-world data and under the erasure channel conditions (Section \ref{errasure_chan}) to a conventional split learning/inference system trained without considering channel effects. %Although this approach might not be optimal in terms of hyperparameter selection, 
We reused the training setup defined in Section \ref{errasure_chan} and detailed in Table \ref{table_training}, with five different models trained using a fixed symbol erasure probability $p_{tr}$ and $H=M=10$, and tested on a range of symbol erasure probabilities 
$p$.  %The main reason for this approach is to ensure fairness in deploying the proposed design in real-world scenarios and to ensure a fair comparison with the general case.
Key observations from the figure indicate that the performance of the COMSPLIT--based system remains consistent with that of the general IoT case (Fig. \ref{Fig_erasure}), revealing its robustness for deployment across different real--world IoT systems. %More precisely, system which design is 
COMSPLIT--based design remains adaptable to a varying channel conditions, as opposed to conventional SL design that experiences significant performance degradation. %Once again, we underline the importance of channel conditions introduced during the training process. 
Note that models with $p_{tr}=0.1-0.2$ (red and green solid lines in Fig. \ref{Fig_erasure_case2}) perform best under good channel conditions  ($p=0-0.2$), whereas models with $p_{tr}=0.4-0.5$ (brown and black solid lines in Fig. \ref{Fig_erasure_case2}) exhibit better performance when channel conditions deteriorate ($p>0.4$). This aligns with the findings from Section \ref{errasure_chan} and highlights the potential for additional system intelligence, where the server can select the appropriate model based on current channel conditions. However, comparing Figs. \ref{Fig_erasure} and \ref{Fig_erasure_case2}, we observe certain degradation in system performance in the real-world use case, possibly due to less consistent quality of data in real-world environmental monitoring deployment.

Fig. \ref{Fig_erasure_case2} (red dotted line) illustrates the performance of the early-exit strategy, which, despite expectedly reduced performance compared to the general IoT system use case, remains effective under poor channel conditions ($p>0.8$). On the other side,  in the AWGN channel scenario, the system's behavior, both with and without the early--exit strategy, aligns with the findings from use case 1 (Section \ref{awgn_perf}). 
%This demonstrates significant robustness to varying channel conditions, though there is a slight performance degradation attributable to lower data quality, as previously discussed for the erasure channel.}

\subsubsection{Heterogeneous IoT Devices}
\begin{figure}[t]
	\begin{tikzpicture}[spy using outlines=
{rectangle, magnification=4, connect spies}]
  	\begin{semilogyaxis}[width=1\columnwidth, height=7.5cm, 
	legend style={at={(0.21,0.98)}, anchor= north,font=\scriptsize, legend style={nodes={scale=0.9, transform shape}}},
   	legend cell align={left},
	legend columns=1,   	 
   	x tick label style={/pgf/number format/.cd,
   	set thousands separator={},fixed},
   	y tick label style={/pgf/number format/.cd,fixed, precision=2, /tikz/.cd},
   	xlabel={$p$},
   	ylabel={MSE},
   	label style={font=\footnotesize},
   	grid=major,   	
   	xmin = 0, xmax = 0.9,
   	ymin=0.008, ymax=0.3,
   	line width=0.8pt,
   	tick label style={font=\footnotesize},]
   	\addplot[blue, mark=square] 
   	table [x={x}, y={y}] {./Figs./Dunav/hetIoT/mse_1_0};
   	\addlegendentry{$MSE_1$-No channel}
        \addplot[dashed, blue, mark=square, mark options=solid] 
   	table [x={x}, y={y}] {./Figs./Dunav/hetIoT/mse_2_0};
   	\addlegendentry{$MSE_2$-No channel}

        \addplot[red, mark=x] 
   	table [x={x}, y={y}] {./Figs./Dunav/hetIoT/mse_1_01};
   	\addlegendentry{$MSE_1$-$p_{tr}=0.1$}
        \addplot[dashed, red, mark=x, mark options=solid] 
   	table [x={x}, y={y}] {./Figs./Dunav/hetIoT/mse_2_01};
   	\addlegendentry{$MSE_2$-$p_tr=0.1$}

        %\addplot[purple, mark=o] 
   	%table [x={x}, y={y}] {./Figs./hetIoT/mse_1_03};
   	%\addlegendentry{$MSE_1$-$p_{tr}=0.3$}
        %\addplot[dashed, purple, mark=o, mark options=solid] 
   	%table [x={x}, y={y}] {./Figs./hetIoT/mse_2_03};
   	%\addlegendentry{$MSE_2$-$p_{tr}=0.3$}

        \addplot[black, mark=star] 
   	table [x={x}, y={y}] {./Figs./Dunav/hetIoT/mse_1_05};
   	\addlegendentry{$MSE_1$-$p_{tr}=0.5$}
        \addplot[dashed, black, mark=star, mark options=solid] 
   	table [x={x}, y={y}] {./Figs./Dunav/hetIoT/mse_2_05};
   	\addlegendentry{$MSE_2$-$p_{tr}=0.5$}
    
 	\end{semilogyaxis}
	\end{tikzpicture}
	\vspace*{-5mm}
	\caption{Heterogeneous IoT Devices: $MSE$ versus symbol erasure probability $p$ performances for 2 devices ($\mathcal{C}(f_{\text{E}_1})<\mathcal{C}(f_{\text{E}_2})$) under different $p_{tr}$ - Use Case 2.}
	\label{Fig_hIoT_2}
\end{figure}
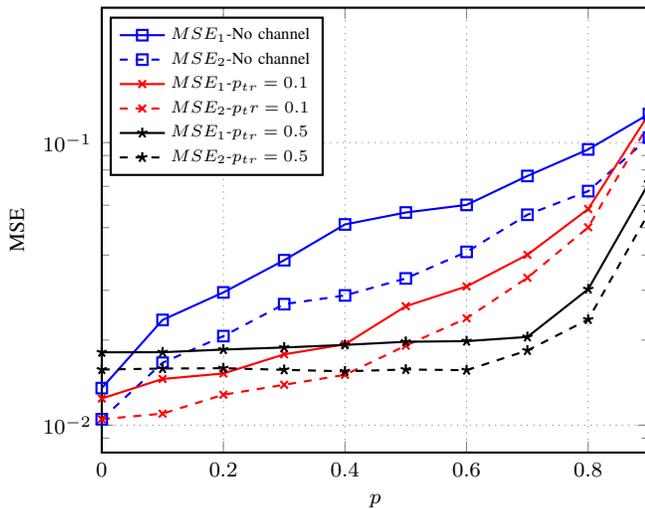

The most complex aspect of the proposed COMSPLIT design involves integrating the wireless channel into the joint training of multiple devices. To evaluate the feasibility of this scenario with real-world data, we reuse the setup introduced in Section \ref{het_IoT_perf} and further detailed in Table \ref{table_training}, which involves two devices with different computational capabilities. The performance results are presented in Fig. \ref{Fig_hIoT_2}, where $MSE_1$ is associated with the edge sub--network $f_{\text{E}_1}$, and $MSE_2$ is associated with $f_{\text{E}_2}$ (Fig. \ref{fig_hIoT}). As observed, the main conclusion from Section \ref{het_IoT_perf} remains valid: the COMSPLIT design significantly outperforms conventional split learning/inference algorithms for $p>0.1$. Additionally, the more capable edge device, which manages two LSTM layers, achieves lower $MSE$ values (illustrated by the dashed lines in Fig. \ref{Fig_hIoT_2}, $MSE_1>MSE_2$), highlighting the advantage of enhanced computational resources. The proposed design is suitable for real-world implementation, as it can leverage multi-modal data for improved predictions and environmental understanding, enabling the server to control computational load of each edge device in the real--world system.

\subsection{COMSPLIT Practical Implementation Challenges}
Communication--aware SL models demonstrate improved inference on both synthetic and real--world data (Sections \ref{results} and \ref{results_dunav}) in theoretical scenarios investigated in this work. However, their deployment in real--world environments will encounter challenges discussed in Section \ref{sec:related}. In our future work, we will explore real-world design and deployment of end-to-end system. In the process, we will investigate various solutions and adapt them to fit the specific context of environmental monitoring systems. For instance, enhancing the adaptability of DL models for heterogeneous IoT platforms (with different computational capabilities) can be achieved by incorporating model pruning and quantization techniques \cite{le_2024}. Resource management in real--world SL deployment can be improved by utilizing cluster--based parallel training/inference, significantly reducing overall system latency \cite{wu_2023}. In the proposed water quality monitoring scenario (Section \ref{scenWater}), IoT devices are integrated into smart buoys, housing all sensors and communication equipment (Fig. \ref{fig_dunav_scen}). Consequently, only a few buoys (down to a single one) are needed at each location, reducing scalability and scheduling concerns for this specific scenario. However, the proposed design can be easily expanded to accommodate various time series--based IoT scenarios, as demonstrated in Section \ref{scen1}. In these cases, scalability might be addressed through the integration of blockchain technology, which enables the creation of serverless distributed learning frameworks and enhances scalability \cite{le_2024}, while scheduling could benefit from incorporating reinforcement learning approaches \cite{agrawal_2020}.   

\section{Conclusion}
This paper explores a novel approach to designing SL paradigm with a focus on communication awareness. We provide a thorough examination of how the proposed SL algorithm performs under diverse channel models and conditions. Our approach involves incorporating an extra channel layer into the SL architecture during the offline training process, thereby bolstering the system's resilience to various distortions commonly encountered in wireless channels in real--world testing phase. Consequently, the proposed approach significantly outperforms basic SL approaches that were constructed and trained without channel integration, for both erasure and AWGN channels. Furthermore, by incorporating an early-exit strategy, there is significant potential to achieve satisfactory performance even in scenarios with severely corrupted channel conditions. This is especially notable when prior knowledge is integrated into the learning process through a weighted loss function. Finally, we introduced a system that is not only aware of communication but also considers the diverse computational capabilities of edge devices within an IoT network, as well as the various modalities of data sources. Such heterogeneous system provide a comprehensive understanding of the network environment, enabling efficient resource allocation and tailored data processing strategies. This work advances the design of the SL paradigm in IoT networks, offering valuable insights into how IoT systems are expected to behave under real-world conditions. These findings pave the way for future research that will focus on deploying the proposed design into real-world IoT systems, including the integration of UAV-assisted communication, especially considering the promising results obtained from a real-world dataset.

\newpage

\vfill

\end{document}